\documentclass[showpacs,aps,prd,reprint,superscriptaddress,nofootinbib,longbibliography]{revtex4-2}

\usepackage[dvipdfmx]{graphicx}
\usepackage{color}
\usepackage{amsmath,amssymb,bm,color,longtable,mathrsfs,amsfonts,slashed,ulem}

\usepackage[colorlinks=true, pdfstartview=FitV, linkcolor=magenta,citecolor=blue, urlcolor=magenta,
bookmarks=true, bookmarksnumbered=true, breaklinks]{hyperref}

\allowdisplaybreaks

\newcommand \beq{\begin{eqnarray}}
\newcommand \eeq{\end{eqnarray}}

\newcommand{\Slash}[1]{{\ooalign{\hfil/\hfil\crcr$#1$}}}

\newcommand{\Nc}{N_{\rm c}}
\newcommand{\Nf}{N_{\rm f}}

\newcommand{\lqcd}{\Lambda_{\rm QCD}}
\newcommand{\vp}{ {\bm p}}

\newcommand{\vK}{{\bm K}}

\newcommand{\vx}{ \bm{ {x}} }

\newcommand{\la}{\langle}
\newcommand{\ra}{\rangle}

\newcommand{\calL}{\mathcal{L}}

\newcommand{\calH}{\mathcal{H}}

\newcommand{\calD}{\mathcal{D}}

\newcommand{\rmd}{\mathrm{d}}
\newcommand{\rmi}{\mathrm{i}}
\newcommand{\rme}{\mathrm{e}}

\newcommand{\msbar}{ \overline{\rm MS} }
\newcommand{\MS}{ \overline{\rm MS} }

\newcommand{\blueflag}[1]{{\color{blue} #1}}

\begin{document}
\begin{flushright}
\end{flushright}

\title{Sound velocity peak and conformality in isospin QCD}

\author{Ryuji Chiba}
\email{ rjchiba@nucl.phys.tohoku.ac.jp }
\affiliation{ Department of Physics, Tohoku University, Sendai 980-8578, Japan}

\author{Toru Kojo}
\email{ toru.kojo.b1@tohoku.ac.jp }
\affiliation{ Department of Physics, Tohoku University, Sendai 980-8578, Japan}

\date{\today}

\begin{abstract}
We study zero temperature equations of state (EOS) in isospin QCD within a quark-meson model which is renormalizable 
and hence eliminates high density artifacts in models with the ultraviolet cutoff (e.g., NJL models).
The model exhibits a crossover transition of pion condensations from the Bose-Einstein-Condensation regime at low density to the Bardeen-Cooper-Schrieffer regime at high density.
The EOS stiffens quickly and approaches the quark matter regime at density significantly less than the density for pions to spatially overlap.
The squared sound velocity $c_s^2$ develops a peak in the crossover region, and
then gradually relaxes to the conformal value $1/3$ from above, in contrast to the perturbative QCD results which predicts the approach from below.
In the context of QCD computations, this opposite trend is in part due to the lack of gluon exchanges in our model,
and also due to the non-perturbative power corrections arising from the condensates.
We argue that with large power corrections the trace anomaly can be negative.
Our EOS reproduces the qualitative trend of the lattice results from the BEC to the BCS regime,
implying that the quark-meson model captures relevant effective degrees of freedom.
The BCS gap in our model is $\Delta \simeq 300$ MeV in the quark matter domain, 
and naive application of the BCS relation for the critical temperature $T_c \simeq 0.57\Delta$ yields the estimate $T_c \simeq 170$ MeV,
in good agreement with the lattice data.

\end{abstract}


\maketitle


\section{Introduction}
\label{sec:Introduction}

The quantum chromodynamics (QCD) with a large isospin chemical potential ($\mu_I$) 
can be studied in lattice Monte-Carlo simulations and hence has been a useful laboratory to test theoretical conceptions 
in dense matter \cite{Son:2000xc,Son:2000by,Splittorff:2000mm}.
In this theory, the positive isospin chemical potential favors the population of up-quarks and of down-antiquarks.
A matter with finite isospin density starts with a Bose-Einstein-Condensation (BEC) phase of charged pions as composite particles.
The dilute regime is well-described by chiral effective theories 
(ChEFT) for pions \cite{Adhikari:2020qda,Adhikari:2020kdn,Adhikari:2020ufo,Adhikari:2019zaj,Andersen:2018qkq}.
As density increases the quark substructure of pions should become important and the system transforms into 
a Bardeen-Cooper-Schrieffer (BCS) phase with a substantial quark Fermi sea.
This BEC-BCS transition is crossover (for BEC-BCS crossover, see, e.g., Ref.~\cite{Leggett_book,schrieffer1999theory,BCS-BEC_Parish}), 
as confirmed by model studies and lattice Monte-Carlo simulations.
We study this crossover in the context of quark-hadron continuity \cite{Schafer:1998ef,Hatsuda:2006ps,Baym:2017whm} or duality \cite{McLerran:2007qj,Fujimoto:2023ioi}, 
which may be also realized in QCD at finite baryon chemical potential ($\mu_B$).

One of fundamental topics in dense QCD is the equation of state (EOS) (Ref.~\cite{Kojo:2020krb} for a short review from the QCD perspective), 
which is a crucial piece to understand the structure of neutron stars (NSs).
Recent analyses of NSs, with similar radii ($\simeq 12.4$ km) for $2.1$ and $1.4$ solar mass NSs \cite{Miller:2021qha,Riley:2021pdl,Raaijmakers:2021uju}, 
and the nuclear physics constraints at nuclear saturation density $n_0$ ($\simeq 0.16\, {\rm fm}^{-3}$),
suggest that the EOS stiffens rapidly (i.e., the pressure $P$ grows rapidly as a function of energy density $\varepsilon$)
from low baryon density to high density, of $n_B = 4$-$7n_0$, which is expected to be realized in the core of massive NSs.
This stiffening accompanies the peak of sound velocity, $c_s = (\partial P/\partial \varepsilon)^{1/2}$;
the $c_s^2$ is $\ll 1$ in the nuclear domain, goes beyond the so-called conformal value $1/3$, and relaxes to $1/3$ in the relativistic limit
where the quark kinetic energy dominates over the interaction.
While NS observations suggest such non-monotonic behaviors of $c_s$, 
it is necessary to understand the mechanisms from the microscopic physics.
The sound velocity peak was first indicated in
phenomenological interpolation of hadronic and quark matter EOS \cite{Masuda:2012kf,Masuda:2012ed,Masuda:2015kha,Masuda:2015wva},
discussed in more general ground based on nuclear physics and NS observations by Ref.~\cite{Bedaque:2014sqa},
and further elucidated in Refs.~\cite{Kojo:2014rca,Kojo:2015fua,Fukushima:2015bda} utilizing the quark degrees of freedom.
Recently more detailed descriptions have been attempted, see Refs.~\cite{McLerran:2018hbz,Jeong:2019lhv,Duarte:2020xsp,Duarte:2020kvi,Duarte:2023cki,Zhao:2020dvu,Cao:2020byn,Margueron:2021dtx,Kojo:2021ugu,Hippert:2021gfs},
but it is difficult to directly test the scenarios.
We use the isospin QCD for which lattice simulations are available, and delineate the behavior of $c_s^2$.


Another interesting question is how the $c_s^2$ approaches the conformal limit, $1/3$.
Perturbative QCD (pQCD) \cite{Freedman:1977gz,Freedman:1976ub,Kurkela:2009gj,Annala:2019puf,Gorda:2021kme,Gorda:2021znl}, 
which is supposed to be valid at $n_B \gtrsim 40n_0$, predicts that the $c_s^2$ approaches $1/3$ from below.
The domain between $n_B\simeq 10n_0$ and $\simeq 40n_0$ has not been explored intensively.
For this regime it is natural to regard quarks as relevant degrees of freedom 
but whose properties may be substantially renormalized by strong interaction effects \cite{Suenaga:2019jjv,Kojo:2021knn,Fujimoto:2020tjc};
if such interaction effects are properly absorbed into effective parameters of quasi-quarks, 
it is possible that
the residual interactions may be treated in the same spirit as in constituent quark models for hadron physics.
If this residual corrections are indeed smaller than the relativistic kinetic energy of quasi-particles, 
the system should show the conformal behavior even before achieving weakly correlated quark matter.
How the matter reaches the conformal regime is intensively discussed 
in recent works \cite{Annala:2019puf,Ma:2019ery,Fujimoto:2022ohj,Marczenko:2022jhl,Blaschke:2022egm,Ivanytskyi:2022bjc,Annala:2023cwx}.

We address the non-monotonic behavior of $c_s^2$ in isospin QCD within a renormalizable quark-meson model.
The properties of the model in isospin QCD have been analyzed in detail by Refs.~\cite{Adhikari:2016eef,Adhikari:2018cea,Ayala:2023cnt}.\footnote{
%
		The renormalization condition described in Ref.~\cite{Ayala:2023cnt} differs from ours and Refs.\cite{Adhikari:2016eef,Adhikari:2018cea};
		the former demands the tree level relations to be satisfied at each chemical potential so that the counter terms vary with the chemical potential.
		In contrast, in this paper the counter terms are completely fixed in vacuum. 
		Because of the difference in the renormalization procedure, 
		the behaviors of condensates and EOS of Ref.~\cite{Ayala:2023cnt} appear to be different from ours; 
		in particular, we find the sound velocity peak while Ref.~\cite{Ayala:2023cnt} did not.
}
We follow their renormalization procedures.
The advantage of using renormalizable effective models over models with a UV cutoff (e.g., the NJL models)
is that one can temper the high density artifacts.
In particular, the BCS type states have a distorted quark occupation probability whose high momentum tail reaches very high momenta,
exceeding the UV cutoff. 
This is in contrast to the ideal gas case with the occupation probability $\theta(p-p_f)$ which discontinuously drops to zero at the Fermi momentum $p_f$  before reaching the UV cutoff.
In fact, NJL studies with BCS states exhibit growing $c_s^2$ toward the high density limit \cite{Baym:2019iky,Kojo:2021wax}.
In the quark-meson model such growing behaviors disappear; the $c_s^2$ relaxes to the conformal value $1/3$, as it should.

While our model predicts $c_s^2 \rightarrow 1/3$, the conformal limit is reached from above, not from below as predicted in pQCD.
The latter is due to the density dependence induced through the running $\alpha_s$.
In the weak coupling limit and at high density, the only relevant scale is $\mu_I$ and the  $c_s^2=1/3$ follows from $\partial (P/\mu_I^4)/\partial \mu_I =0$.
The first important corrections to the conformal limit come from the $\lqcd \simeq $ 200-300 MeV in the running coupling constant. 
If we take into account $\lqcd$ only in this way, the $c_s^2$ is reduced from $1/3$.
Meanwhile, at the energy scale around $\sim 1\, {\rm GeV}$, it has been long known that power corrections of $\lqcd$, 
which can not be expressed as perturbative series in $\alpha_s$, play important roles to capture 
the qualitative features in QCD \cite{Novikov:1977dq,Shifman:1978bx,Shifman:1978by}.
Parametrizing pressure with power corrections as ($\chi_I$: isospin susceptibility)
\beq
P(\mu_I) = a_0 \mu_I^4 + a_2 \mu_I^2 \,,
\label{eq:pressure_intro}
\eeq
where $a_2 \sim \lqcd^2$, the squared sound velocity can be expressed as 
\beq
c_s^2 = \frac{\, n_I \,}{\, \mu_I \chi_I \,} = \frac{\, 2a_0 \mu_I^2 + a_2 \,}{\, 6a_0 \mu_I^2 + a_2 \,} \,.
\label{eq:cs2_intro}
\eeq
For a positive $a_2$, the $c_s^2$ is larger than 1/3, and close to 1 if the $a_2$ term dominates.
In our quark-meson model the $a_2$ is related to the pion condensates.
We quantify the relation within our quark-meson model.

We briefly address the trace anomaly in dense matter which measures the breaking of the scale invariance \cite{Ma:2019ery,Fujimoto:2022ohj,Marczenko:2022jhl}.
We argue that changes from the non-perturbative to perturbative vacua add positive contributions to the trace anomaly,
while the power corrections with $a_2 >0$ favors the negative value.
For large power corrections the trace anomaly can be negative.
In this respect the sign of the trace anomaly is very useful to characterize the non-perturbative effects in dense quark matter.

For quantitative aspects, 
we confront our model calculations with the lattice results from two groups \cite{Brandt:2022hwy,Abbott:2023coj}.
Ref.~\cite{Brandt:2022hwy} have more focus on the BEC regime while Ref.~\cite{Abbott:2023coj} covers more global nature up to the pQCD regime.
Both groups agree in the BEC regime, and our model results are consistent with the lattice results.
At high density, our model captures the overall trend of Ref.~\cite{Abbott:2023coj},
especially the sound velocity peak and negative trace anomaly.

Since we are not sure about the convergence of loop expansion,
as supplement studies we perform several parametric studies of EOS to examine several qualitative effects
which we believe to be important.
They are used to delineate the results of the quark-meson model in Sec.\ref{sec:discussion}.

In this paper we use nuclear saturation density in QCD, $n_0 = 0.16\, {\rm fm}^{-3}$, as our unit for isospin density.
While there is no need to address nuclear saturation in isospin QCD,
our goal is to discuss physics as a step to understand NS EOS
and $n_0$ is useful in this phenomenological context.

This paper is organized as follows.
In Sec.\ref{sec:model} we discuss our set up for a quark meson model. The renormalization procedures are summarized.
In Sec.\ref{sec:eos} we present the renormalized thermodynamic potential and resulting EOS, as well as the correlation between condensates and EOS.
We emphasize the importance of quark substructure which can be seen only after including quark loops. 
The numerical results are confronted with the lattice data.
In Sec.\ref{sec:discussion} we discuss the zero point energy in EOS which often appears as the bag constant in a phenomenological model.
In our quark-meson model this quantity can be computed explicitly.
In addition we discuss the power correction to the pQCD.
The evolution of the sound velocity at high density is presented.
We also discuss the trace anomaly as an indicator of the non-perturbative effects.
Section \ref{sec:summary} is devoted to a summary.

\section{Model}
\label{sec:model}

The Lagrangian of the two-flavor quark-meson model is
\begin{align}
	\calL &= {1\over2}\left(\partial_\mu \vec\phi\right)^2 
	- \frac{\, m^2 \,}{2} \vec\phi^2 -{\lambda\over 24}(\vec\phi^2)^2 + h\sigma
	\notag\\
	&- \overline{\psi} \big[ \Slash{\partial} - g(\sigma + \rmi \gamma^5\vec\tau\cdot \vec\pi) \big] \psi \,,
\end{align}
where $\psi$ is a quark field with up- and down-quark components, $\psi = (u,d)^T$.
The $\vec\phi = (\sigma,\vec\pi)$ are meson fields which correspond to the isospin $\bf 1$ and $\bf 3$ representations.
The $\tau_i$'s are the Pauli matrices in flavor space.

We compute the thermodynamic potential at finite isospin density $n_I$, utilizing the isospin chemical potential 
$\mu_I = \mu_u = - \mu_d$  as a Lagrange multiplier.
To correctly identify the corresponding Lagrangian, we should begin with the hamiltonian formalism.
The thermodynamic potential is
\beq
	\Omega = H - \mu_I N_I\,,~~~~~~~~ N_I = \int_{\vx} n_I \,.
\eeq
The isospin density in terms of field variables can be identified by the Noether theorem.
Meson and quark fields transform under isospin transformations as
\beq
	\pi_a \mapsto \exp( \rmi \theta_i T_i)^{\rm adj}_{ab}\pi_b \,,~~~~~ \psi \mapsto  \rme^{ \rmi \theta_i \tau_i} \psi \,,
\eeq
and corresponding conserved current can be written as
\beq
	j^\mu_a = \epsilon_{abc}\pi_b\partial^\mu\pi_c + \delta^{\mu 0} \delta_{a3}\bar{\psi} \gamma^0 \tau_3 \psi \,,
\eeq
where 
the $\epsilon_{abc}$ is the complete anti-symmetric tensor with $\epsilon_{123}=1$.
The isospin density is now 
\beq
n_I = j^0_{a=3}
 = \pi_+ \rmi \partial^0 \pi_- 
-  \pi_- \rmi \partial^0 \pi_+
+ \bar{\psi} \gamma^0 \tau_3 \psi  \,.
\eeq
Writing fields collectively as $\Phi = (\vec{\phi}, \psi)$,
the partition function for $\Phi$ is
\begin{align}
	Z = \int\calD \Pi_\Phi \calD \Phi \exp\left[ \rmi \int_x\left(\dot{\Phi}\cdot \Pi_\Phi - \calH + \mu_I n_I \right)\right].
\end{align}
where $(\Pi_\Phi)_i = \partial^0 \Phi_i$ is a field conjugate to $\Phi_i$. 
Keeping in mind that $n_I$ contains the conjugate fields $\Pi_{\pi_\pm} = \partial^0 \pi_\pm$, we integrate $\vec{\Pi}_\phi$ to get
\beq
	Z = \int\calD \Phi\exp\left(\rmi \int_x \calL_B \right) \,.
\eeq
%
Here the Lagrangian at finite density is
\begin{align}
	\calL_B (\vec\phi_B)=&\, {1\over2}\left[(\partial_\mu \sigma_B)^2+(\partial_\mu \pi_{B3} )^2\right] \notag\\
		&\hspace{-1.2cm} + (\partial_\mu + 2 \rmi \mu_I\delta_\mu^0) \pi_B^+\left(\partial^\mu - 2 \rmi \mu_I\delta^\mu_0\right)\pi_B^- \notag\\
		&\hspace{-1.2cm} -{\, m_B^2 \, \over 2} \vec\phi_B^2  -{\lambda_B \over 24} ( \vec\phi_B^2 )^2  + h_B \sigma_B  \notag\\
		&\hspace{-1.2cm} + \overline{\psi}_B \left[ \rmi \Slash{\partial} + \mu_I \tau_3 \gamma^0 - g_B (\sigma_B + \rmi \gamma^5 \vec\tau \cdot \vec\pi_B) \right]\psi_B \,.		
\end{align}
where we attached the subscript $B$ to emphasize that the parameters and fields are unrenormalized.

Below we construct a one-loop effective potential 
within the leading order of the $1/\Nc$ expansion.
In this approximation, meson loop effects on quarks are neglected,
while quark loop effects play crucial roles 
in renormalizing meson parameters as well as the amplitude of meson condensates.
This quark substructure affects the density evolution of meson condensates
and hence the EOS.

First we rewrite the Lagrangian using the renormalized parameters and fields.
We begin with the $O(4)$ symmetric scheme and later relate those renormalized parameters to those in the on-shell scheme.
The bare parameters are written with $O(4)$ symmetric renormalized fields and couplings as
\beq
\phi_B &=& Z_{\phi}^{1/2} \phi\,, ~~~~~~\psi_B = Z_{\psi}^{1/2}  \psi \,, \notag \\
g_B &=& \tilde{Z}_g Z_\psi^{-1} Z_{\phi}^{-1/2} g = Z_g g\,, \notag \\
m_{ B}^2 &=& \tilde{Z}_{m^2} Z_\phi^{-1} m^2 =  Z_{m^2} m^2\,, \notag \\
\lambda_{ B} &=& \tilde{Z}_\lambda Z_\phi^{-2} \lambda = Z_\lambda  \lambda\,, \notag \\
h_B &=& \tilde{Z}_h Z_\phi^{-1/2} h = Z_h h \,.
\label{eq:reno_para}
\eeq
We also define $\delta Z_i = Z_i - 1$ for $i=\phi, \psi, g, \cdots$ and so on.
The $\tilde{Z}_i$ represents the radiative corrections without those for the external lines.
In our model,
the loop corrections to the quark self-energies and quark-meson vertices
appear only through meson-loops and hence
\beq
 Z_\psi = 1 + O(1/\Nc)\,,~~~ \tilde{Z}_g = 1 + O(1/\Nc)\,.
\eeq
Meanwhile, the meson self-energies and tadpole contain quark loops of $O(\Nc)$
which are combined with $g^2 \sim 1/\Nc$ vertices to yield
\beq
&& Z_\phi = 1 + O( g^2 \Nc ) \,,~~~ \tilde{Z}_{m^2} = 1 + O( g^2 \Nc ) \,, \notag \\
&& \tilde{Z}_h  = 1 + O(g^2\Nc) \,,~~~ \tilde{Z}_\lambda = 1 + O(g^4\Nc/\lambda)\,.
\eeq
and hence one must keep the corrections.
It is useful to note that the relation
\beq
g_B \phi_B = g \phi \,,
~~~~~
Z_g = Z_\phi^{-1/2} \,,
\label{eq:g_and_Z}
\eeq
in the large $\Nc$ limit.
The first relation means that the dynamically generated quark mass and gap are renormalization group (RG) invariant.
The second relation tells that we need to study the meson propagators
to describe the running of $g^2$.

Now the Lagrangian is decomposed into the renormalized part and counter terms as
\beq
\calL_B (\vec\phi_B) = \calL (\vec\phi) + \calL_{\rm c.t.} (\vec \phi)\,,
\eeq
Here $\calL$ is the renormalized Lagrangian where all subscripts $B$ are omitted from $\calL_B$ and couplings are replaced with the renormalized couplings.
The counter terms necessary in the large $\Nc$ limit is
\begin{align}
\hspace{-0.1cm}
\calL_{\rm c.t.} (\vec\phi) 
&= {\delta Z_{\phi} \over2} \big[\, (\partial_\mu\sigma)^2+  (\partial_\mu\pi_3)^2 \,\big]  \notag\\
	& + \delta Z_{\phi} (\partial_\mu + 2 \rmi \mu_I\delta_\mu^0)\pi^+\left(\partial^\mu - 2 \rmi \mu_I\delta^\mu_0\right)\pi^- \notag\\
	& - \frac{\, \delta \tilde{m}^2 \,}{\, 2 \,}  \vec{\phi}^2  
		-{\,  \delta \tilde{\lambda} \, \over 24} (\vec{\phi}^2 )^2 
		+ \delta \tilde{h} \sigma \,.
\label{eq:ct_before}		
\end{align}
The counter Lagrangian is used when we calculate loop corrections.

We construct the effective potential $\Gamma(\phi_0)$ with the $\MS$ normalization of fields.
The effective potentials defined at different renormalization schemes are related as 
$\Gamma_R (\phi_R) = \Gamma_R ( Z_{R'} \phi_{R'}/Z_R ) = \Gamma_{R'} (\phi_{R'})$
where, in $\Gamma_{R'}$, the parameters are replaced as $(g_R, \lambda_R, \cdots) \rightarrow (g_{R'}, \lambda_{R'}, \cdots) $
while the kinetic terms are always normalized to $1$.
Actually, it is more convenient to work with a $\Gamma ( g \phi_0)$
in which $g \phi_0$ is the RG invariant  in the large $\Nc$ limit.
We specify $M_q = g \sigma_0 $ and $\Delta = g (\pi_1)_0$ as variables for the effective potential
and then $\Gamma_R (M_q, \Delta) = \Gamma_{R'} ( M_q, \Delta )$,
i.e., we need only to take into account changes in $(g, \lambda,, \cdots)$
when we change the renormalization conditions.

\subsection{Parameter fixing with vacuum quantities}

\subsubsection{Renormalization of effective potenital}

Now we fix the counter terms by renormalizing physical parameters in vacuum.
The simplest scheme to obtain the renormalized effective potential is the $\MS$ scheme.
The effective potential takes the form
\beq
V_{\rm 1-loop}
&=& 
 \frac{\, m^2 \,}{\, 2 g^2 \,} M_q^2 
 + {\, \lambda \, \over 24 g^4 } M_q^4 
		-  \frac{\, h \,}{\, g \,} M_q  \notag \\
&+&  \frac{\, \delta \tilde{m}^2 \,}{\, 2 g^2 \,}  M_q^2 
+{\,  \delta \tilde{ \lambda} \, \over 24 g^4 } M_q^4 
		-  \frac{\, \delta \tilde{h} \,}{\, g \,} M_q 
 \notag \\
 &+& V_q \,.
\eeq
The $V_q$ is the one-loop contributions from the quark energy
\beq
	V_q = - 2N_c \Nf \int_{\vp} E_D (\vp) \,,~~~E_D = \sqrt{\vp^2 + M_q^2}\,,
\eeq
where we write $M_q = g \sigma_0$.
We will treat these divergences by dimensional regularization $d \to 3-2\epsilon$,
\beq
	\int_{\vp} = \left({e^{\gamma_E}\Lambda^2 \over 4\pi}\right)^\epsilon \int {d^dp \over (2\pi)^d}.
\eeq
where $\Lambda$ is the renormalizing scale introduced by the $\overline{\rm MS}$ scheme and $\gamma_E = 0.577...$ is the Euler-Mascheroni constant. 
The quark energy now reads 
\beq
	V_q &=& {4N_c \over (4\pi)^2} \left({e^{\gamma_E}\Lambda^2 \over M_q^2 }\right)^\epsilon 
	\Gamma(-2+\epsilon) \, M_q^4 \notag \\
	&=&  {2N_c \over (4\pi)^2} \bigg(\frac{1}{\epsilon}  + \frac{3}{2} -  \ln \frac{\, M_q^2 \,}{\, \Lambda^2 \,} \bigg) M_q^4 + O (\epsilon)\,.
\eeq
There is no $1/\epsilon$ pole in the linear and quadratic terms, 
while the $1/\epsilon$ in the quartic term is cancelled by $\delta \tilde{\lambda}$,
\beq
\delta \tilde{h} = 0\,,~~~ \delta \tilde{m}^2 = 0\,,~~~
 \delta \tilde{\lambda} = - \frac{\, 48 \Nc g^4 \,}{\, (4\pi)^2 \epsilon \,}  \,.
\eeq
The effective potential in vacuum now reads
\beq
V_{\rm 1-loop}
&=& 
 \frac{\, m^2 \,}{\, 2 g^2 \,} M_q^2 
 + \frac{\, \lambda \,}{\, 24 g^4 \,} M_q^4 
		- \frac{\, h \,}{\, g \,} M_q 
 \notag \\
 &+& 
  \frac{\, 2N_c \,}{\,  (4\pi)^2 \,} \bigg( \frac{3}{2} -  \ln \frac{\, M_q^2 \,}{\, \Lambda^2 \,} \bigg) M_q^4 
  \,.
\eeq
We demand the effective potential to be RG invariant, i.e.,
the effective potential does not change by replacement of $\Lambda \rightarrow \Lambda'$.
This must be valid for a given $M_q$, so each coefficient of $M_q$ must be invariant.
The invariance of $M_q$ and $M_q^2$ terms 
requires $(m/g, h/g )$ do not run in the large $\Nc$ limit,
as consistent with Eq.~\eqref{eq:reno_para}.
Meanwhile the $M_q^4$ terms contain the $\ln \Lambda^2$ factor so that
\beq
\frac{\partial}{\, \partial \ln \Lambda^2 \,} \bigg( \frac{\, \lambda (\Lambda) \,}{\, g^4(\Lambda) \,} \bigg)
 = - \frac{\, 48N_c \,}{\,  (4\pi)^2 \,}  \,,
\eeq
or
\beq
\frac{\, \partial \ln \lambda \,}{\, \partial \ln \Lambda^2 \,}
- 2 \frac{\, \partial \ln g^2 \,}{\, \partial \ln \Lambda^2 \,}
= - \frac{\, 48N_c g^4 \,}{\,  (4\pi)^2 \lambda \,}  \,,
\label{eq:RG_lambda}
\eeq
The running of $g^2$ is obtained from the analyses of field normalizations of $\phi$,
thanks to the relation $g^2 (\Lambda) = g_B^2/ Z_\phi (\Lambda)$.

\subsubsection{Renormalization of meson propagators}

Now we consider the renormalization conditions for mesons to fix the $Z_\phi$.
We write $M_0$ as a solution to minimize the effective potential and use it to compute the meson self-energies.
We demand that the $\sigma$ and $\pi$ have the pole at $p^2 =m_\sigma^2$ and $p^2 = m_\pi^2$, respectively,
%
\beq
&&m^2 + \frac{\, \lambda \,}{\, 2g^2 \,} M_0^2 
+ \Sigma_{\sigma} (m_\sigma^2) 
= m_\sigma^2 \,, 
\notag \\
&&m^2 + \frac{\, \lambda \,}{\, 6 g^2 \,} M_0^2 
+ \Sigma_{\pi} (m_\pi^2) 
= m_\pi^2 \,.
\eeq
%
The self-energies include the quark one-loop contributions $\Sigma^q_{\sigma, \pi}$ and counter terms
\beq
&& \Sigma_{\sigma} (m_\sigma^2)
= \Sigma^q_{\sigma} (m_\sigma^2) - \delta Z_\phi m_\sigma^2 + \frac{\, \delta \tilde{\lambda} \,}{\, 2 g^2 \,} M_0^2 
\,, \notag \\
&& \Sigma_{\pi} (m_\pi^2)
= \Sigma^q_{\pi} (m_\pi^2) - \delta Z_\phi m_\pi^2 + \frac{\, \delta \tilde{\lambda} \,}{\, 6 g^2 \,} M_0^2 \,.
\eeq
The quark loop $\Sigma^q_{\sigma, \pi}$ is UV divergent.
The counter term $\delta \tilde{\lambda}$ automatically cancels the UV divergences coupled to $M_0$. 
The $\delta Z_\phi$ is arranged to cancel $m_\sigma$- and $m_\pi$- dependent UV divergences in $\Sigma^q_{\sigma, \pi}$, 
\beq
\delta Z_\phi = - \frac{\, 4 g^2 \Nc  \,}{\, (4\pi)^2 \epsilon \,} 
\,,
\eeq
with which 
\beq
\Sigma_\sigma (p^2)
&=&
\frac{\, 8 g^2 \Nc  \,}{\, (4\pi)^2 \,}
 \bigg[  
  M_0^2 
  - G_\sigma 	
  + \frac{\, 6 M_0^2 - p^2 \,}{2} 
	  \ln \frac{\, \Lambda^2 \,}{\, M_0^2  \,} 
 \bigg] \,,
 \notag \\
 \Sigma_\pi  (p^2)
&=&
\frac{\, 8 g^2 \Nc  \,}{\, (4\pi)^2 \,}
 \bigg[  
  M_0^2 
 - G_\pi 
  + \frac{\, 2 M_0^2 - p^2 \,}{2} 
	  \ln \frac{\, \Lambda^2 \,}{\, M_0^2  \,} 
 \bigg] \,,
 \notag \\
 \label{eq:meson_self}
\eeq
where $G_\sigma $ and $G_\pi $ are functions of $p^2$ and $M_0^2$,
\beq
&&G_\sigma (p^2) = \frac{\, p^2 - 4M_0^2 \,}{2} F (p^2)\,,
~~
G_\pi (p^2) = \frac{\, p^2  \,}{2} F (p^2)\,, \notag \\
&&
~~~ F(p^2) = - \int_0^1 \rmd x~ \ln \bigg( 1 - \frac{\, p^2 x(1-x) \,}{\, M_0^2 \,} \bigg) \,.
\eeq
It is useful to note $G_\sigma (4M_0^2) = 0$ and $G_\pi (0) = 0$.
Later we also make use of $F(0) = 0$ and $F(4M_0^2) = -2$.

We note that the parameter $\Lambda$ manifestly appears in the self-energies
but the pole positions should be RG invariant.
This demands
\beq
\frac{\, \partial \, \delta Z_\phi (\Lambda) \,}{\, \partial \ln \Lambda^2 \,}  
\simeq
\frac{\, \partial \ln Z_\phi (\Lambda) \,}{\, \partial \ln \Lambda^2 \,}  
=
- \frac{\, 4 g^2 \Nc  \,}{\, (4\pi)^2 \,}  \,.
\eeq
Eqs.\eqref{eq:g_and_Z} and \eqref{eq:RG_lambda} lead to
\beq
 \frac{\, \partial g^2 \,}{\, \partial \ln \Lambda^2 \,}
 = \frac{\, 4 g^4 \Nc  \,}{\, (4\pi)^2 \,}  \,,
 ~~~
 \frac{\, \partial \lambda \,}{\, \partial \ln \Lambda^2 \,}
= \frac{\, 8N_c g^2 ( \lambda - 6 g^2 ) \,}{\,  (4\pi)^2  \,}  \,.
\notag \\
\eeq
The effective potential and the pole locations with running parameters are RG invariant,
so below we choose $\Lambda = M_0$ to get rid of the $\ln \Lambda$ terms.

Finally we also mention how the $\MS$ and on-shell renormalization schemes are related.
Here we discuss only $Z_\sigma^{\rm OS}$ as we will fix $f_\pi$ by $\la \sigma_{\rm OS} \ra = f_\pi$.
We note
\beq
\la \sigma \sigma \ra 
= \frac{\, \big( 1 - \frac{\, \rmd \Sigma_\sigma \,}{\, \rmd p^2 \,} \big|_{p^2=m_\sigma^2} \big)^{-1} \,}{\, p^2 - m_\sigma^2 \,}
= \frac{\, Z_\sigma^{\rm OS} \,}{\, Z_\phi \,} \la \sigma \sigma \ra_{\rm OS}
\eeq
where the residue of $\la \phi \phi \ra_{\rm OS}$ is normalized to 1.
Thus 
\beq
\hspace{-0.5cm}
 &&\frac{\, Z_\phi \,}{\, Z_\sigma^{\rm OS} \,} - 1
 =  \frac{\, 8 g^2 \Nc  \,}{\, (4\pi)^2 \,} \frac{\, \rmd G_\sigma \,}{\, \rmd p^2 \,} \big|_{p^2=m_\sigma^2} 
 \notag \\
 && 
 \hspace{0.5cm}
 =  \frac{\, 4 g^2 \Nc  \,}{\, (4\pi)^2 \,}
 \bigg( F(m_\sigma^2) + (m_\sigma^2 - 4M_0^2 ) F'(m_\sigma^2) \bigg)
 \,.
\eeq
In the parameter range of our interest, we find $Z_\phi < Z_\sigma^{\rm OS}$.
For instance, for theories with $m_\sigma=2M_0$, 
the inequality is verified by noting $F(4M_0^2) = -2$.

\subsubsection{Parameter fixing}

We fix the values of parameters in our model.
To evaluate the effective potential,
we need to fix four parameters $(m, g, \lambda, h)$ at $\Lambda = M_0$.
Our input is $(f_\pi, M_0, m_\sigma, m_\pi)$.

First we fix the value of $g$. We note that $M_0$ is RG invariant (in the large $\Nc$ limit),
\beq
M_0 = g \sigma_0 = g_{\rm OS} f_\pi \,.
\eeq
We can fix $g_{\rm OS} = M_0 /f_\pi$ while $g$ can be fixed by the relation 
\beq
g^2  
= \bigg( \frac{\, Z_{g}^{\rm OS}\,}{\, Z_g \,} \bigg)^2 g_{\rm OS}^2
=  \frac{\, Z_\phi \,} {\, Z_{\sigma}^{\rm OS}\,} g_{\rm OS}^2
~~\big( < g_{\rm OS}^2 \big)
 \,.
\eeq
For typical parameter set $M_0 \sim 300$ MeV and $f_\pi \sim 90$ MeV, $g_{\rm OS} \sim 3.3$ which is large.
In the $\MS$ scheme $g^2$ is smaller and the expansion of $g^2$ is slightly better in systematics.

Having $g$ fixed, we can determine $m^2$ and $\lambda$ from the pole conditions for $m_\sigma $ and $m_\pi$,
\beq
m^2 
&=& 
- \frac{\, \big( m_\sigma^2 - 3 m_\pi^2 \big) - \big( \Sigma_\sigma(m_\sigma^2)- 3 \Sigma_\pi ( m_\pi^2) \big) \,}{2} \,, 
\notag \\
\lambda 
&=& 3g^2 \frac{\,  \big(  m_\sigma^2 - m_\pi^2 \big) 
- \big( \Sigma_\sigma(m_\sigma^2)-  \Sigma_\pi ( m_\pi^2) \big)  \,}{\, M_0^2 \, } \,.
\eeq
To get analytic insights, it is again useful to consider the case $m_\sigma = 2M_0$ and $m_\pi =0$.
Then 
\beq
\hspace{-0.5cm}
\lambda \rightarrow 12 g^2 \,,~~~
m^2 
\rightarrow 
- 2M_0^2
- \frac{\, 8 g^2 \Nc \,}{\, (4\pi)^2 \,} M_0^2\,.
\eeq
In this limit, it is clear that $m^2$, which drives the $\sigma$ condensation at tree level, becomes more negative than the tree level counterpart by radiative corrections.
This limit also suggests that $\lambda$ is typically large, of $O$(10-100).

Finally we fix $h$. Using the parameters defined at $\Lambda = M_0$, the effective potential takes the form
\beq
V_{\rm 1-loop}
&=& 
 \frac{\, m^2 \,}{\, 2 g^2 \,} M_q^2 
 + \frac{\, \lambda \,}{\, 24 g^4 \,} M_q^4 
		- h \frac{\, M_q \,}{\, g \,}  
 \notag \\
 &+& 
  \frac{\, 2N_c \,}{\,  (4\pi)^2 \,} \bigg( \frac{3}{2} -  \ln \frac{\, M_q^2 \,}{\, M_0^2 \,} \bigg) M_q^4 
  \,.
\eeq
%
%
%
%
%
The gap equation at $M_q=M_0$ fixes the value of $h$,
\beq
\hspace{-0.5cm}
h
= \frac{\, M_0 \,}{g} \bigg(
m^2  + \frac{\, \lambda \,}{\, 6 g^2 \,} M_0^2 + \frac{\, 8 \Nc \,}{\, (4\pi)^2 \,} M_0^2  
\bigg)
\,.
\eeq
Using the condition for the pion pole, one can rewrite it as
\beq
h = m_\pi^2 \bigg( 1 + \frac{\, 4 g^2 \Nc  \,}{\, (4\pi)^2 \,} F (m_\pi^2) \bigg) \frac{\, M_0 \,}{g}
= \frac{\, Z_h^{\rm OS} \,}{\, Z_h \,} h^{\rm OS}   \,,
\eeq
where $h^{\rm OS} = m_\pi^2 f_\pi$,
the standard expression in the chiral EFT.
%

\subsection{At finite isospin density}

For a large isospin chemical potential, either $\pi_1$ or $\pi_2$ can condense while $\pi_3$ fields are unaffected.
Without loss of generality we assume the $\pi_1$ to condense.
The quark part in the unperturbed Lagrangian acquires an extra term
\beq
	\calL_2^{\rm extra} = \overline{\psi} \rmi\gamma^5 \tau_1 \Delta \psi \,,
\eeq
with which the quark propagator becomes the BCS type propagator. 
The poles exist at
\beq
	E_u = E_{\overline{d}} = E(\mu_I)\,,~~~ E_d=E_{\overline{u}}=E(-\mu_I) \,,
\eeq
where (see the derivation in appendix \ref{appsec:derivation})
\beq
	E(\mu_I) = \sqrt{\left(E_D-\mu_I\right)^2+\Delta^2} \,.
\eeq
The $u$ and $\bar{d}$ quark excitations cost at least the energies of the BCS gap $\sim \Delta$.
Meanwhile $d$ and $\bar{u}$ quarks need large energies of $\sim M+\mu_I$ to get excited.

The effective potential in the $\MS$ scheme is
\beq
V_{\rm 1-loop}
&=& 
 \frac{\, m^2 \,}{\, 2 g^2 \,} \big( M_q^2 + \Delta^2 \big) 
 + {\, \lambda \, \over 24 g^4 } \big( M_q^2 + \Delta^2 \big)^2  
		-  \frac{\, h \,}{\, g \,} M_q  \notag \\
&-&  \frac{\, 2 \mu_I^2 \,}{\, g^2 \,} \big( 1 + \delta Z_\phi \big) \Delta^2 
+ {\,  \delta \tilde{ \lambda} \, \over 24 g^4 } \big( M_q^2 + \Delta^2 \big)^2
 \notag \\
 &+& V_q (\mu_I, M_q, \Delta)  \,.
\eeq
We note that the $\mu_I$ dependent term contains a UV divergent counter term
which is necessary to cancel a $\mu_I$ dependent UV divergence from $V_q$.

The single particle energies depend on $M_q$ and $\Delta$ in medium,
\beq
	V_q = -N_c\int_{\vp} \big( E_u+E_d+E_{\overline{u}}+E_{\overline{d}} \big) \,.
\eeq
To get analytic insights, we split
\beq
V_q  = V_q^R +  V_q^{(0)} + V_q^{(2)} \,,
\eeq
where the upper script specifies the power of $\mu_I$,
\beq
V_q^{(0)} &=& - 4\Nc \int_{\vp} \sqrt{ E_D^2+\Delta^2} \,, 
\notag \\
V_q^{(2)} &=& - 2 \Nc \int_{\vp} \frac{\, \mu_I^2 \Delta^2 \,}{\, ( E_D^2+\Delta^2)^{3/2} \,} \,, 
\eeq
whose computations can be carried out with the dimensional regularization,
\beq
V_q^{(0)} 	&=&  {2N_c \over (4\pi)^2} \bigg(\frac{1}{\epsilon}  + \frac{3}{2} -  \ln \frac{\, M_q^2 + \Delta^2 \,}{\, \Lambda^2 \,} \bigg) \big( M_q^2 + \Delta^2 \big)^4 \,,
\notag \\
V_q^{(2)} 	&=&  {4N_c \over (4\pi)^2} \bigg(\frac{1}{\epsilon}  -  \ln \frac{\, M_q^2 + \Delta^2 \,}{\, \Lambda^2 \,} \bigg) \big( - 2\mu_I^2 \Delta^2 \big)  \,.
\eeq
We have extracted up to $\mu_I^2$ terms as they contain the UV divergences,
while $V_q^R$ is UV finite and contains terms which scale as $\mu_I^4$ and vanish at $\mu_I=0$. 
At large $\mu_I$, $V_q^R$ dominates over the other terms as far as $\Delta$ and $M_q$ do not grow as $\sim \mu_I$.
We numerically evaluate $V_q^R$ and found that
$V_q^R \simeq b_0 \mu_I^4 + b_2 \mu_I^2 + \cdots$ 
with $b_0 \simeq - \Nc \mu_I^4/6\pi^2$ and $b_2 \simeq 0$.
Hence, the $\mu_I^2$ components of the effective potential 
are well-saturated by the $V_q^{(2)}$.

The effective potential in the $\MS$ scheme now reads
\beq
V_{\rm 1-loop}
&=& 
 \frac{\, m^2 \,}{\, 2 g^2 \,} \big( M_q^2 + \Delta^2 \big) 
 + {\, \lambda \, \over 24 g^4 } \big( M_q^2 + \Delta^2 \big)^2  
		-  \frac{\, h \,}{\, g \,} M_q  
 \notag \\
 &+& 
  \frac{\, 2N_c \,}{\,  (4\pi)^2 \,} \bigg( \frac{3}{2} -  \ln \frac{\, M_q^2 + \Delta^2 \,}{\, M_0^2 \,} \bigg) \big( M_q^2 + \Delta^2 \big)^2 
\notag \\
&-& 2 \mu_I^2 \bigg( \frac{\, 1 \,}{\, g^2 \,}  - \frac{\, 4 \Nc  \,}{\, (4\pi)^2 \,} \ln \frac{\, M_q^2 + \Delta^2 \,}{\, M_0^2 \,} \bigg) \Delta^2
 \notag \\
 &+& V_q^R (\mu_I, M_q, \Delta)  \,.
\eeq
The effective potential rewritten with hadronic parameters is shown in Appendix.~\ref{appsec:QM_pressure}
and we use it to evaluate $M_q$ and $\Delta$, as was done in Refs.~\cite{Adhikari:2016eef,Adhikari:2018cea}.
The expectation value $M_{q*}$ and $\Delta_*$ are determined by the gap equations,
\beq
	\left.\frac{\, \partial V_{\rm 1-loop} \,}{\partial M_q }\right|_{M_{q*},\Delta_*} = 0\,, ~~
	\left.\frac{\, \partial V_{\rm 1-loop} \,}{\partial \Delta}\right|_{M_{q*},\Delta_*} = 0 \,.
\eeq
In the next section we examine the behaviors of condensates 
and the relation to the thermodynamics.

\section{Equations of state}
\label{sec:eos}

We now numerically examine the mean field EOS.
Unless otherwise stated, 
we fix the model parameters to satisfy the following vacuum parameters\footnote{
Here we have used the sigma mass as the renormalization condition
but in reality the sigma or $f_0(500)$ state has a broad width.
This width has been studied and confirmed in the linear sigma model, 
which is very similar to this model, by Ref.~\cite{Jido:2000} considering the $\sigma\to\pi\pi$ scattering process.
In our study at large $\Nc$, quark loops enter only condensed mesons and counter terms for mesons,
but do not affect mesonic fluctuations or meson excitations,
and hence the impacts of meson width are not addressed.
}:
\beq
	m_\pi &=& 140\, {\rm MeV}\,,~~~~~ m_\sigma = 600\, {\rm MeV}\,, \notag\\
	f_\pi &=& 90\, {\rm MeV} \,,~~~~~~\, M_0 = 300\, {\rm MeV} \,.
\eeq
which correspond to the following {\it on-shell} coupling constants,
$g_{\rm OS} \simeq 3.33$ and $\lambda_{\rm OS} \simeq 126$.\footnote{ 
In the $\MS$ scheme, the couplings are smaller.
The details depend on the choice of $m_\sigma$
which is more uncertain than the other input parameters.
For $m_\sigma = 2M_0$,  
there is simple relation
\beq
\hspace{-0.5cm}
g^2 &=& \frac{\, g_{\rm OS}^2 \,}{\, 1 + \frac{\, 8g_{\rm OS}^2 \,}{\, (4\pi)^2 \,} \,} 
\,,~~
\notag \\
\lambda 
&=& 12g^2 \bigg[1 - \frac{\, m_\pi^2 \,}{\, 4M_0^2 \,} \bigg(1+\frac{\, 4g^2 \Nc F (m_\pi^2) \,}{\, (4\pi)^2 \, } \bigg) \bigg] \,,
\eeq
from which $g \simeq 2.03$.
This reduces the value of $\lambda$ by a factor $ \sim (g/g_{\rm OS} )^2 \simeq 0.37$.
The value of $\lambda$ becomes even smaller for $m_\pi \rightarrow m_\sigma$.
}

The large couplings in the present one-loop analyses are worrisome.
Meanwhile it has been known that constituent quark type models with couplings of $O(1)$ 
work remarkably well without rigorous justifications.
In this work we simply hope that the similar situation holds in our studies.
We also note that, 
in the case of the nucleon-meson model, whose structure is very similar to the quark-meson model, 
the tree and one-loop results are qualitatively different, 
but the difference between one-loop results and the functional renormalization group results
are quantitative one, the order of $\sim 30\%$ \cite{Brandes:2021pti}.
Thus we expect our one-loop results to be useful to gain some qualitative insights 
into the overall trend of isospin QCD.

With this qualification in mind, we proceed to the examination of the EOS.
For comparison to the lattice data in Ref.~\cite{Abbott:2023coj}, later we also examine the $m_\pi = 170$ MeV case 
with $(m_\sigma, f_\pi, M_0)$ kept the same as the $m_\pi=140$ MeV case.

\subsection{Evolution of microscopic quantities}
\label{sec:evo_micro}

\begin{figure}[tbph]
\centering
\vspace{-0.5cm}
\includegraphics[width=1.05\linewidth]{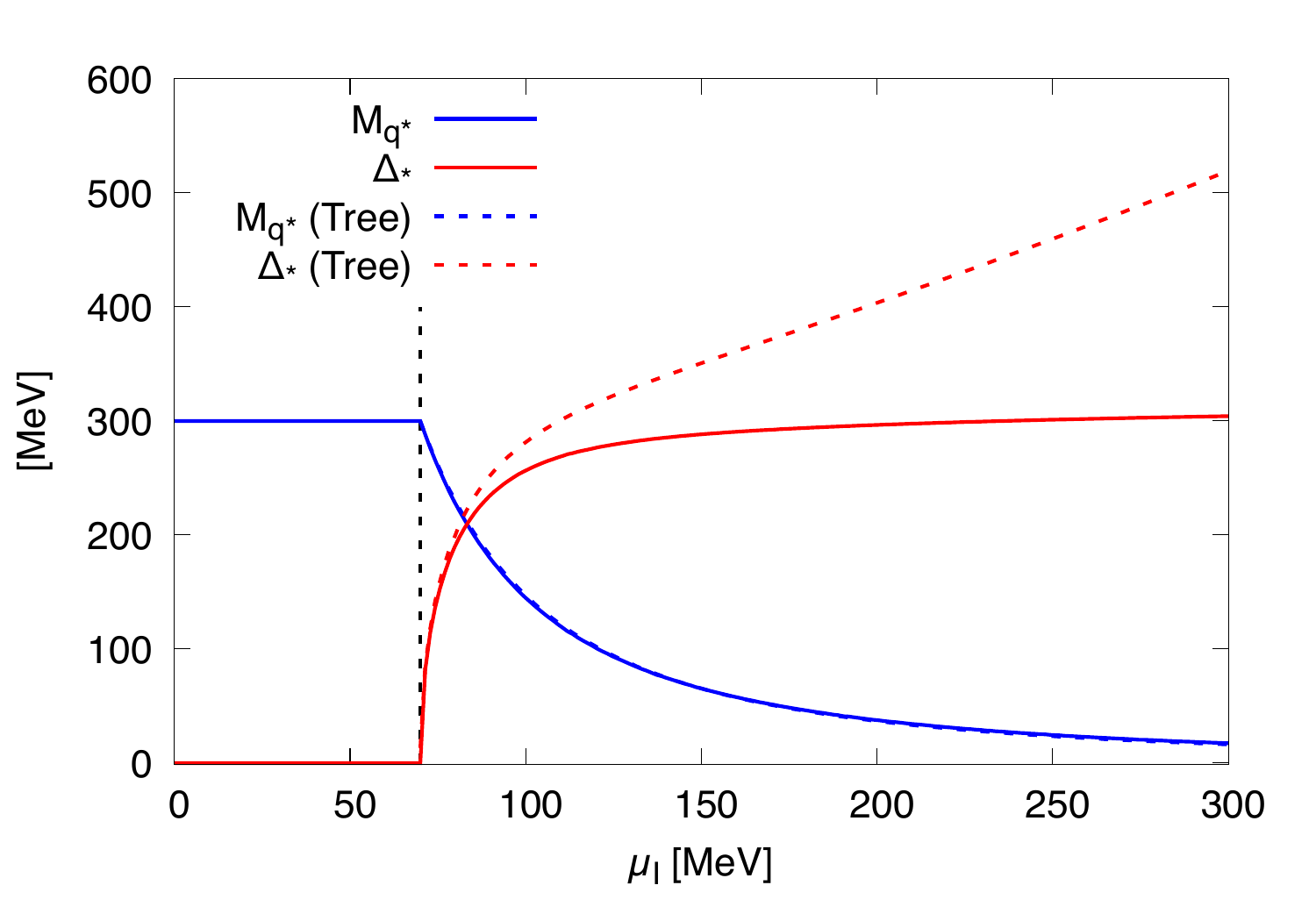}
\caption{
	Chiral and pion condensates as functions of a scaled chemical potential $\mu_I/m_\pi$. }
\label{fig:condensate}
\end{figure}

\begin{figure}[tbph]
\centering
\vspace{-1.2cm}
\includegraphics[width=1.0\columnwidth]{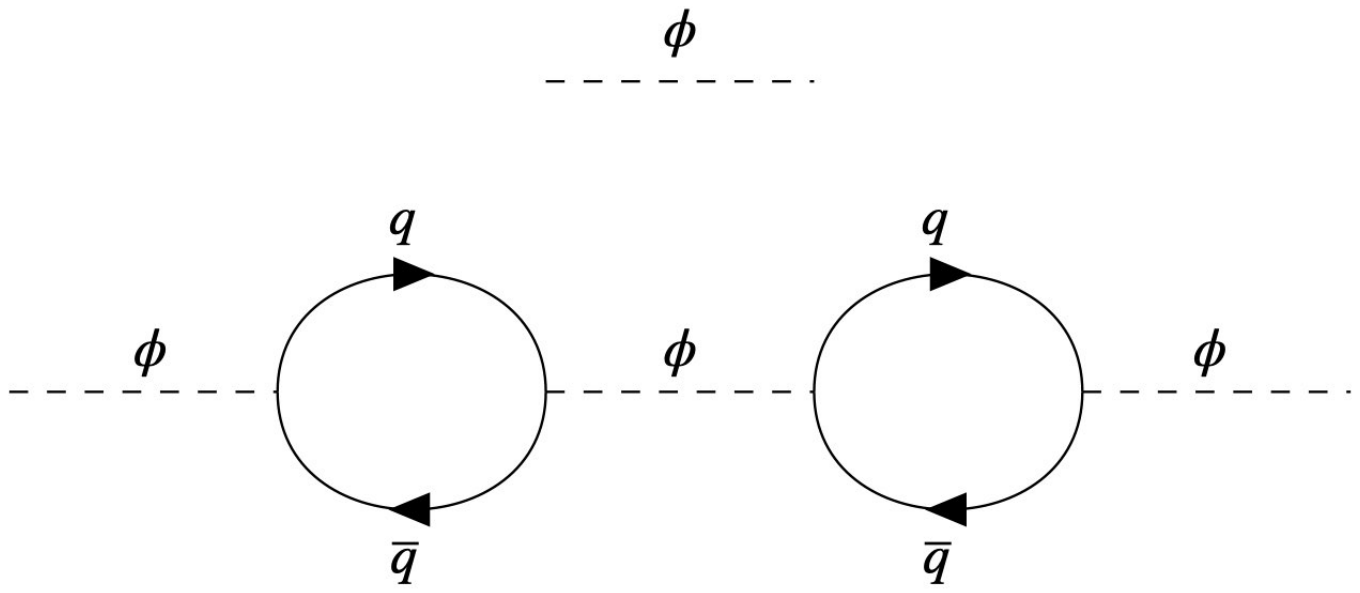}
\vspace{-2.cm}
\caption{
	{\bf Top}: A meson propagator as an elementary particle. 
	{\bf Bottom}: A meson propagator with a quark loop. The meson can be interpreted as a composite particle. 
	}
\label{fig:mesonpropagator}
\end{figure}

First we examine how condensates evolve as functions of $\mu_I$.
Shown in Fig.\ref{fig:condensate} are the constituent quark mass and the gap associated with the pion condensate.
For $M_q$, there are no significant differences between the tree-level (dashed blue line) and one-loop (solid blue line) results.
Meanwhile, the pion condensate $\Delta$ at tree level increases linearly with $\mu_I$, whereas, at one-loop, it converges to a finite value, 
$ \Delta \simeq 300\, {\rm MeV}$.
This drastic change of behavior indicates that the one-loop correction has more physical contents than mere perturbative corrections.

At tree-level, the Lagrangian makes no reference to quarks so that the mesons are treated as elementary particles (Fig.\ref{fig:mesonpropagator} ({\bf Top})).
By adding quark loops, however, they no longer can be regarded as purely elementary particles.
If we regard mesons as fundamental, quark loops are regarded as corrections to the meson dynamics.  
But if we regard quark descriptions as more fundamental, 
mesons are intermediate states appearing in the quark-antiquark scattering processes (Fig.\ref{fig:mesonpropagator} ({\bf Bottom})).

In this study, we keep only the leading $\Nc$ contributions and hence
the quark substructure effects on meson {\it fluctuations} are not reflected in EOS (as meson loops are suppressed by $1/\Nc$).
However, the quark substructure effects do affect {\it condensed} mesons by tempering the amplitudes.
The quark loops change the structure of the present theory 
and in this sense it may not be appropriate to call quark loops as corrections;
rather they should be regarded as leading order contributions.

Now we examine how quark loops qualitatively change the behavior of $\Delta$.
To address this question we work with the $\MS$ expression for the moment as it takes a concise form.
We consider a large $\mu_I$ and assume $M_q \ll \Delta$.
At tree level, the effective potential behaves as
\beq
	V_{\rm tree}  ~\rightarrow~ - \frac{\, 2\mu_I^2 \,}{g^2} \Delta^2 + {\lambda \over 24 g^4} \Delta^4 \,,	
	\label{eq:tree_dense}		
\eeq
and the solution of the gap equation is found by balancing $\mu^2 \Delta^2$ and $\Delta^4$ terms.
Hence $\Delta_{\rm tree} \sim \mu_I$ inevitably follows.
Note also that $\lambda >0$, like the hard core repulsion, plays an essential role to stop the growth of $\Delta$.

Including quark loops, however, the coefficient of $\mu_I^2$ term acquires the $ \ln (\Delta/M_0)^2$ term
which, before $\Delta^4$ becomes dominant, can stop the growth of $\Delta$.
At large $\mu_I$ and assuming $\Delta \ll \mu_I$,
\beq
\hspace{-0.5cm}
V_{\rm 1-loop} 
\simeq
- 2 \mu_I^2 \bigg( \frac{\, 1 \,}{\, g^2 \,}  - \frac{\, 4 \Nc  \,}{\, (4\pi)^2 \,} \ln \frac{\, \Delta^2 \,}{\, M_0^2 \,} \bigg) \Delta^2
+ V_q^R \,,
\label{eq:V_formal}
\eeq
where $V_q^R \sim \mu_I^4$ weakly depends upon $\Delta$.
Then the gap equation is determined by the coefficient of $\mu_I^2$ term,
\beq
\hspace{-0.5cm}
\frac{\, \partial V_{\rm 1-loop} \,}{\, \partial \Delta \,} 
\simeq
- \frac{\, 8 \mu_I^2 \Nc\,}{\, (4\pi)^2 \,}\bigg( \frac{\, (4\pi)^2 \,}{\, 4 g^2 \Nc \,}  - 1 - \ln \frac{\, \Delta^2 \,}{\, M_0^2 \,} \bigg) \simeq 0 \,.
\eeq
The solution is $\mu_I$-independent,\footnote{
The $g^2$-dependence in this expression may be confusing and here we give supplemental comments.
At large $g \gg 1$, $\Delta^2 \rightarrow M_0^2 \rme^{-1} $ which looks smaller than $M_0^2$.
This reduction is fictitious; if we hold $(m, \lambda, h)$ fixed and increase $g$, 
$M_0 \sim g f_\pi$ increases and $\Delta$ also increases.
On the other hand, at small $g \ll 1$, apparently $\Delta$ becomes larger,
but our assumption of $\mu_I \gg \Delta$ and hence our estimate is violated.
In this situation, the size of $\Delta$ is primarily determined by the tree level relation
as the hadronic and quark sectors decouple for $g\rightarrow 0$.
}
\beq
\Delta_*^2 \simeq M_0^2 \, \rme^{ \frac{\, (4\pi)^2 \,}{\, 4 g^2 \Nc \,}  - 1 } \,.
\eeq
For our parameter set, $ (4\pi)^2/4 g^2 \Nc \sim 1 $ and the exponent is small; 
we find $\Delta \simeq M_0$ as shown in Fig.\ref{fig:condensate}.

Substituting the solution into Eq.\eqref{eq:V_formal}, the $1/g^2$ and the logarithmic terms cancel,
leaving the $- \mu_I^2 \Delta_*^2$ term.
As a result the pressure $P (\mu_I) =-V_{\rm 1-loop} (M_q^*, \Delta_*;\mu_I)$ has the $+ \mu_I^2 \Delta_*^2$ term,
\beq
\hspace{-0.5cm}
P (\mu_I)
\simeq
  \frac{\, \Nc  \,}{\, 2 \pi^2 \,} \mu_I^2 \Delta_*^2
+ P_q^R (\mu_I) \,,
\label{eq:power_from_gap}
\eeq
where $P_q^R = - V_q^R (M_q^*, \Delta_*)$.
The $\mu_I^2$ dependence can be interpreted as the Fermi surface effects with the phase space $\sim 4 \pi p_F^2$
with $p_F$ being the quark Fermi momentum $p_F \sim \mu_I$.

In this expression for large $\mu_I$, {\it hadronic parameters disappear}.
The hadronic parameters $m^2$ and $\lambda$ are neglected because they appear as 
$m^2 \Delta^2$ and $\lambda \Delta^4$ terms much smaller than $\mu_I^2 \Delta^2$ and $\mu_I^4$,
while $g^2$ is absorbed into the expression of $\Delta_*$.
The resulting expression can be most naturally understood in terms of quarks
with non-perturbative effects near the Fermi surface whose strength depends upon the hadron physics.

\subsection{Equations of state}
\label{sec:eos_sub}

%
%
%

Starting with the thermodyamic pressure $P (\mu_I) =-V_{\rm 1-loop} (M_q^*, \Delta_*;\mu_I)$,
the isospin and energy densities are given by
\beq
n_I = \frac{\, \partial P \,}{\, \partial \mu_I \,} \,,~~~~~ \varepsilon = \mu_I n_I - P \,.
\eeq
We study the sound velocity
\beq
c_s^2 = \frac{\, \partial P \,}{\, \partial \varepsilon \,} = \frac{\, n_I \,}{\, \mu_I \chi_I \,}\,,~~~~~
\chi_I = \frac{\, \partial^2 P \,}{\, \partial \mu_I^2 \,} \,,
\eeq
where $\chi_I$ is the isospin susceptibility.

In the following we compare our results with the lattice data 
in Refs.~\cite{Brandt:2022hwy} and \cite{Abbott:2023coj}.
The setup of the former is $\Nf =2+1$ flavors of rooted staggered quarks with the quark masses at the physical point.
The pion decay constant is $f_\pi \simeq 92$-96 MeV for the lattice spacing explored
(the definition of $f_\pi$ differs by a factor $\sqrt{2}$ from ours and we have corrected it).
It should be noted that 
their results at $T=0$ are obtained by correcting the data at small but finite $T$ using the ChEFT.
Beyond $\mu_I \gtrsim m_\pi$ or $n_I \gtrsim 0.5 n_0$ 
the lattice data is not available in Ref.~\cite{Brandt:2022hwy}.
Meanwhile, the lattice data in Ref.~\cite{Abbott:2023coj} using $m_\pi \simeq 170$ MeV and a different formalism
is more suitable to explore high density region up to $ \mu_I \sim 7.5 m_\pi \simeq 1.3$ GeV
(our definition of $\mu_I$ is a half of that in Ref.~\cite{Abbott:2023coj}, taken into account in our figures).

\begin{figure}[tbp]
\vspace{-0.2cm}
\centering
\includegraphics[width=1.0\linewidth]{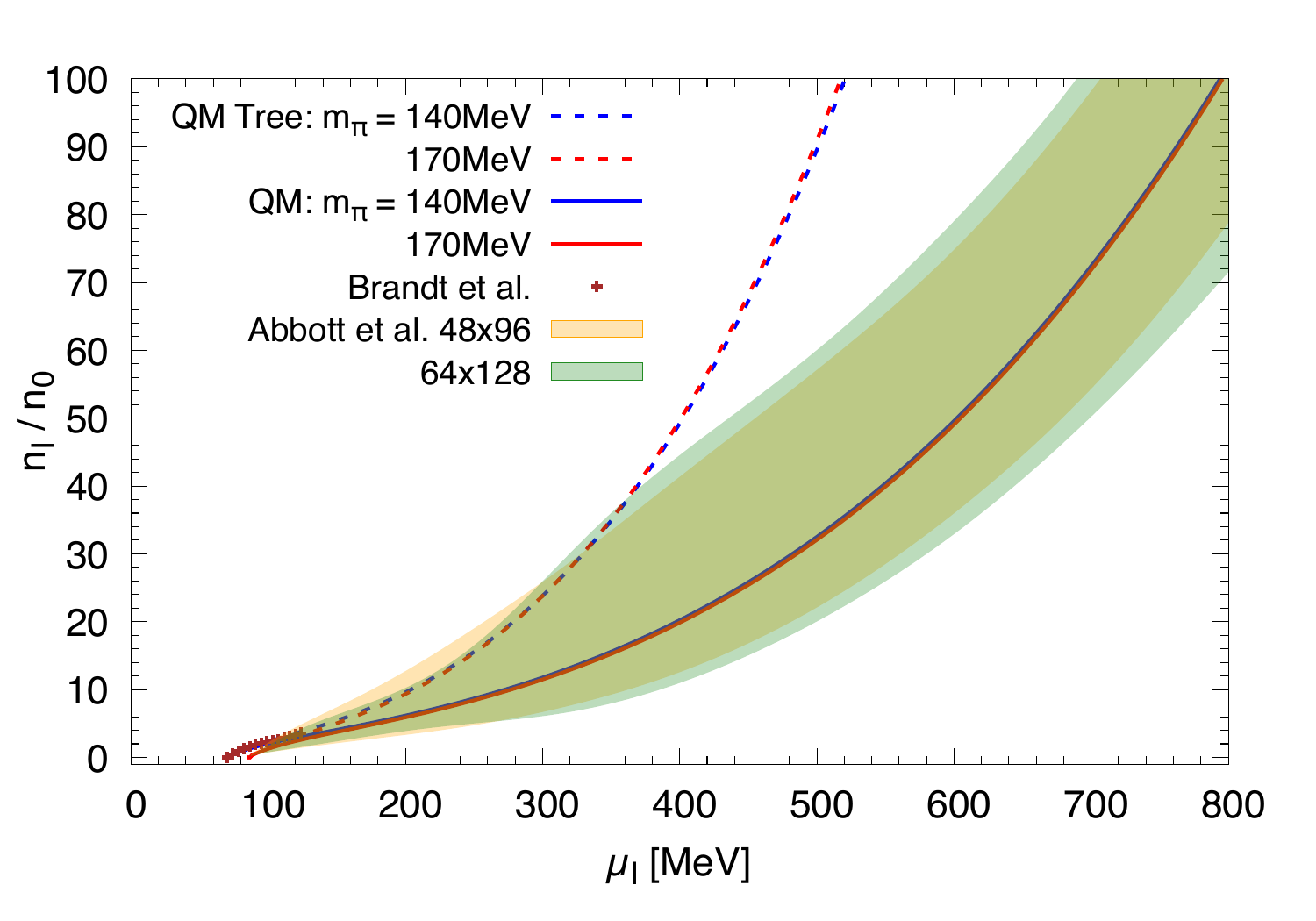}
\includegraphics[width=1.0\linewidth]{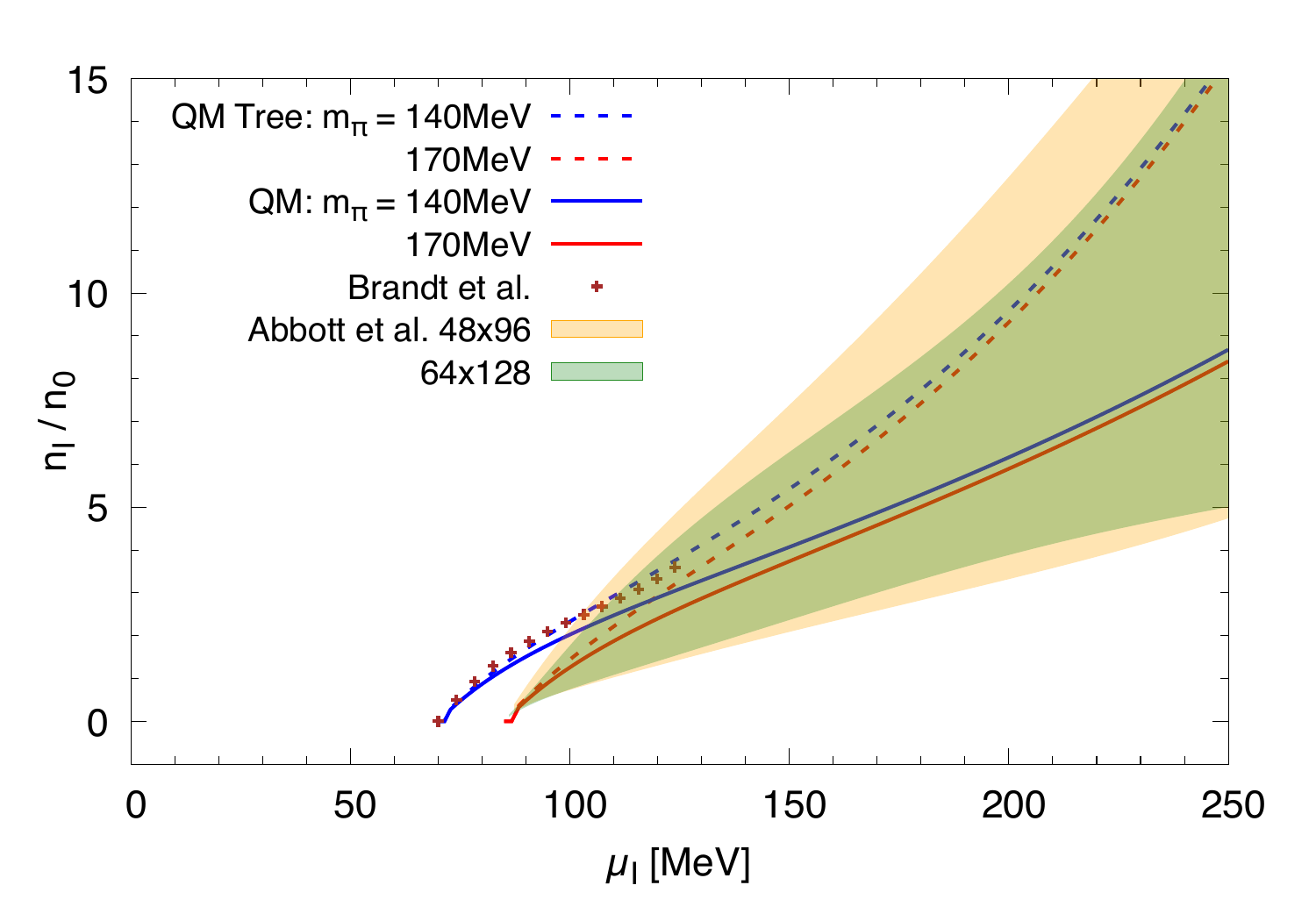}
\vspace{-0.8cm}
\caption{
Isospin density $n_I$ as a function of $\mu_I$,
for the global (upper) and low density (lower) behaviors.
The data points at low density are from Brandt et al. \cite{Brandt:2022hwy} (available up to $\mu_I \simeq 0.9m_\pi$ with $m_\pi \simeq 140$ MeV) 
and bands are from Abbott et al. \cite{Abbott:2023coj} with $m_\pi \simeq 170$ MeV. 
(The definition of $\mu_I$ in Ref.~\cite{Abbott:2023coj} differs from ours by a factor 2 and this is taken into account in our figures.)
}	
\label{fig:n_vs_mu}
\end{figure}
\begin{figure}[tbph]
\vspace{-0.3cm}
\centering
\includegraphics[width=1.0\linewidth]{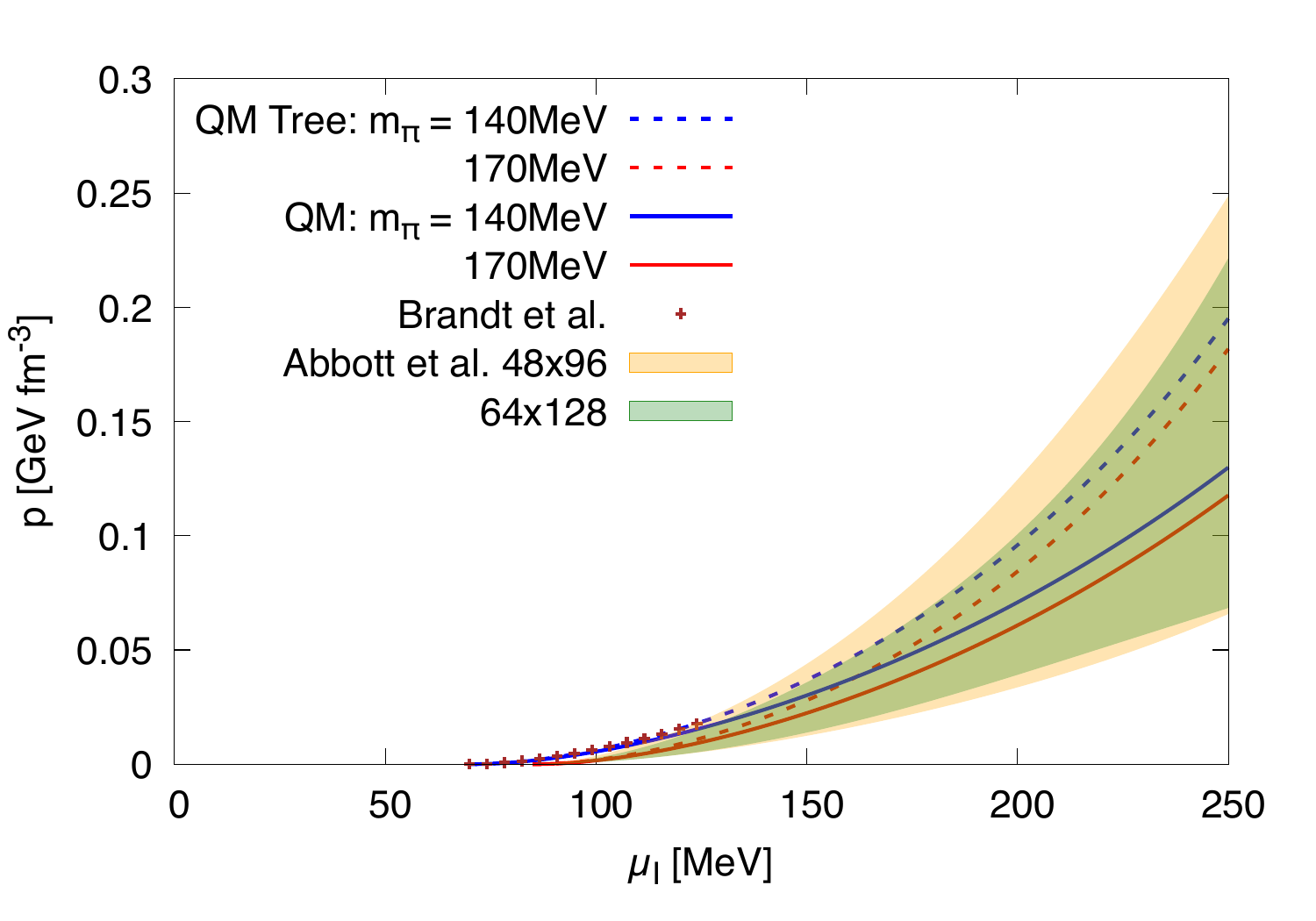}
\includegraphics[width=1.0\linewidth]{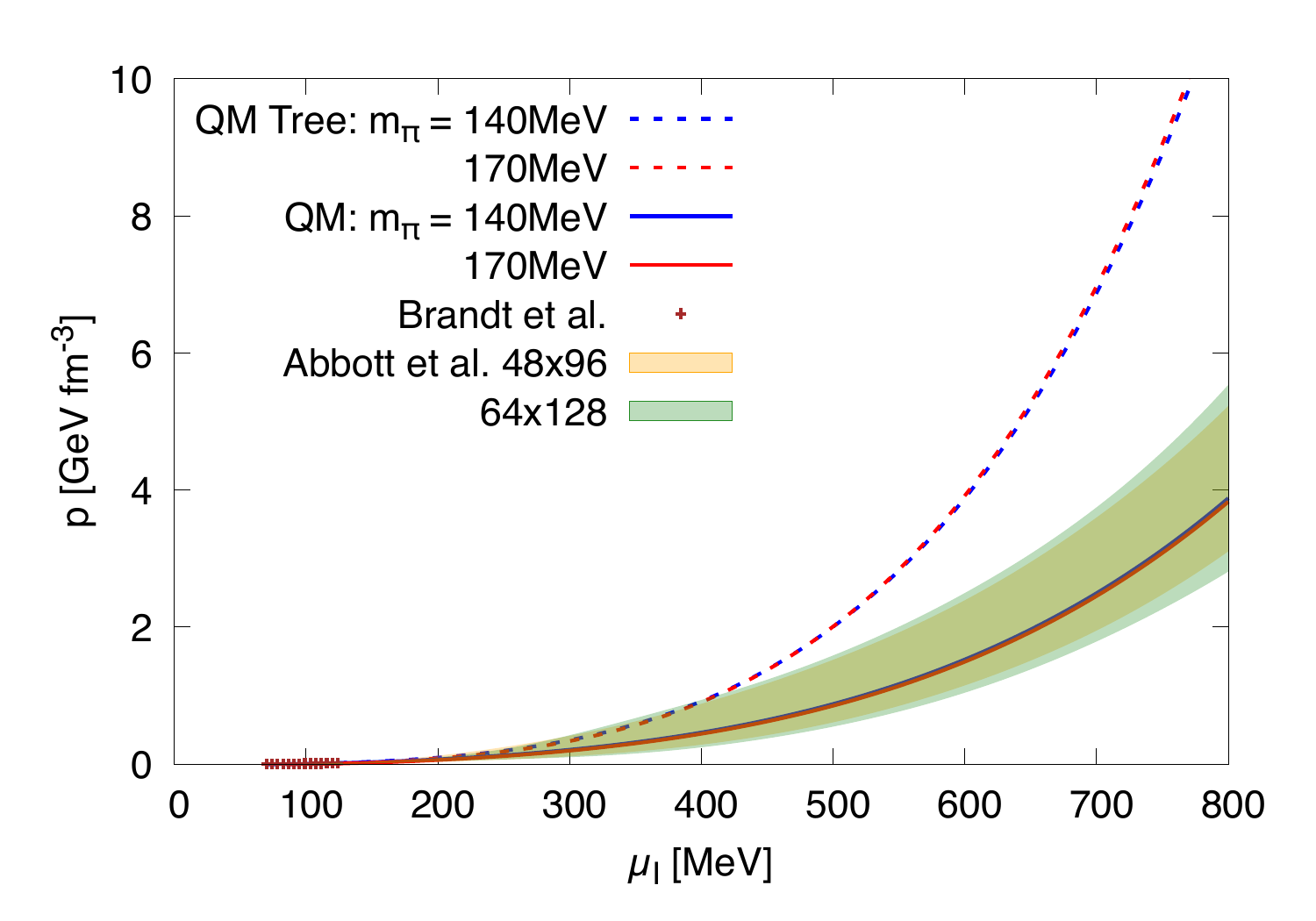}
\vspace{-0.8cm}
\caption{
Pressure $P$ as a function of $\mu_I$ for the low density (upper panel) and global (lower panel) behaviors.
}
\label{fig:p_vs_mu}
\end{figure}
\begin{figure}[htbp]
\vspace{-0.3cm}
\centering
\includegraphics[width=1.0\linewidth]{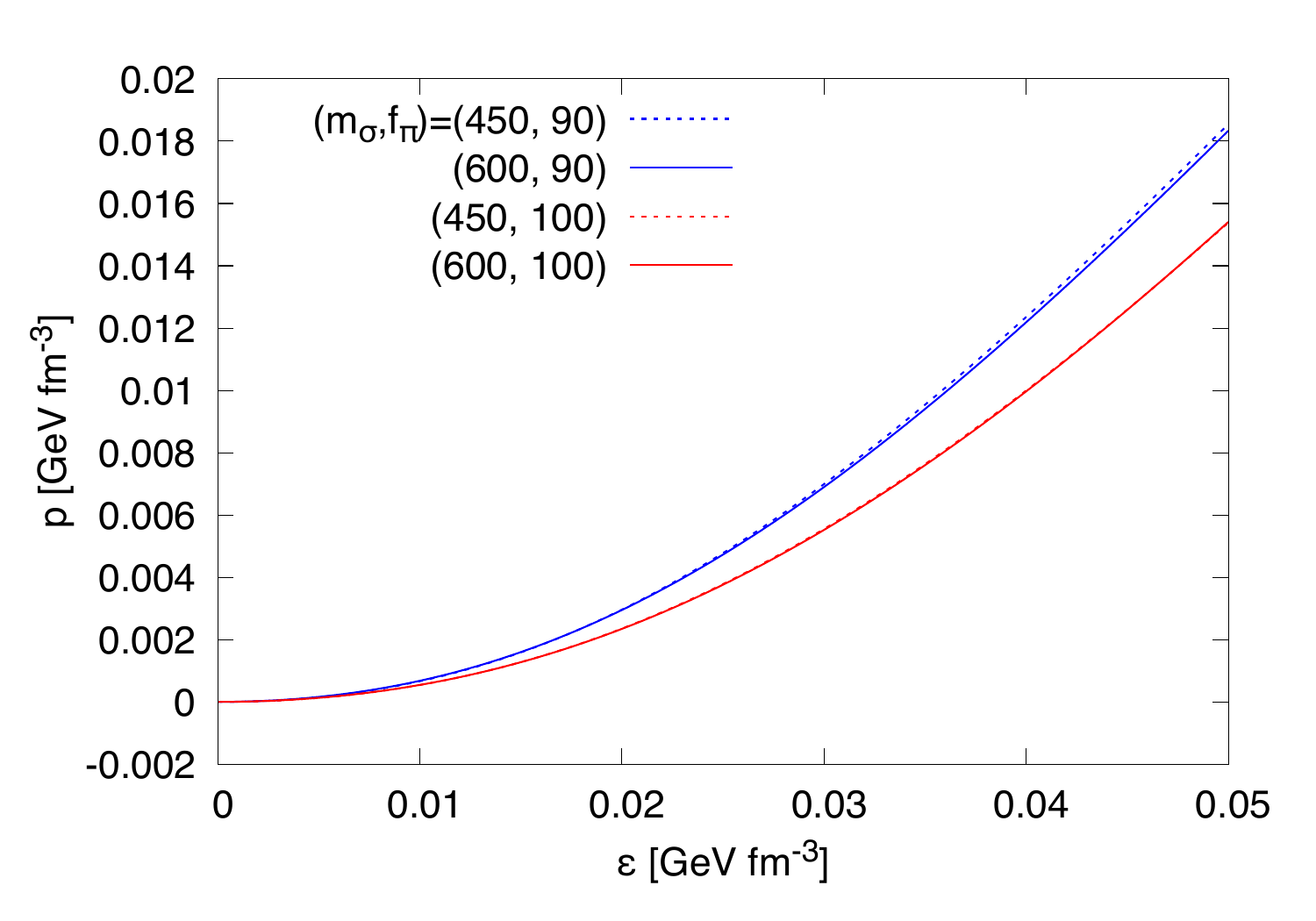}
\vspace{-0.8cm}
\caption{
	Pressure vs energy density for different $(m_\sigma, f_\pi)$ in the MeV unit.
	Larger chiral symmetry breaking, i.e., larger $m_\sigma$ and $f_\pi$, reduces the pressure at a given energy density.
	}
\label{fig:p_vs_e_fpi_dep}
\end{figure}

Figure \ref{fig:n_vs_mu} shows the isospin density $n_I$ as a function of the isospin chemical potential $\mu_I$,
for the global (upper panel) and low density (lower panel) behaviors.
We use $(m_\sigma, f_\pi) = (600, 90)$ MeV and consider $m_\pi =140$ and $170$ MeV for comparison with the lattice data of Refs.~\cite{Brandt:2022hwy} and \cite{Abbott:2023coj}.
The upper panel in Fig.~\ref{fig:n_vs_mu} is specialized for the examination of global features to high density. 
As expected from the qualitative difference in the behavior of condensates, the tree and one-loop results become very different toward high density.
In purely hadronic descriptions, we found $P \sim \lambda \Delta_{\rm tree}^4 \sim \lambda \mu_I^4$ 
whose asymptotic behavior is controlled by the hadronic coupling $\lambda$, the strength of ``hard core repulsion'' between mesons.
This scaling behaviors are changed by quark loops, 
with which the scaling $P \sim c \mu_I^4$ is controlled by the phase space factor for quarks, rather than parameters for hadronic interactions.

In the lower panel of Fig.~\ref{fig:n_vs_mu}, we make more detailed comparison at low density.
 We note that the ChEFT results of Ref.~\cite{Adhikari:2019zaj} including chiral loop corrections agree well with the lattice results of Ref.~\cite{Brandt:2022hwy} (see Fig.2 in Ref.~\cite{Adhikari:2019zaj} for various comparisons),
 while our model results slightly underestimate $n_I$ for a larger $\mu_I$.
 We note that the chiral loops in the ChEFT and quark-loops cover different types of quantum fluctuations.
In Fig.~\ref{fig:p_vs_mu}, we also show the relation between $P$ and $\mu_I$
for the low density and global behaviors.

\begin{figure}[htbp]
\vspace{-0.3cm}
\centering
\includegraphics[width=1.0\linewidth]{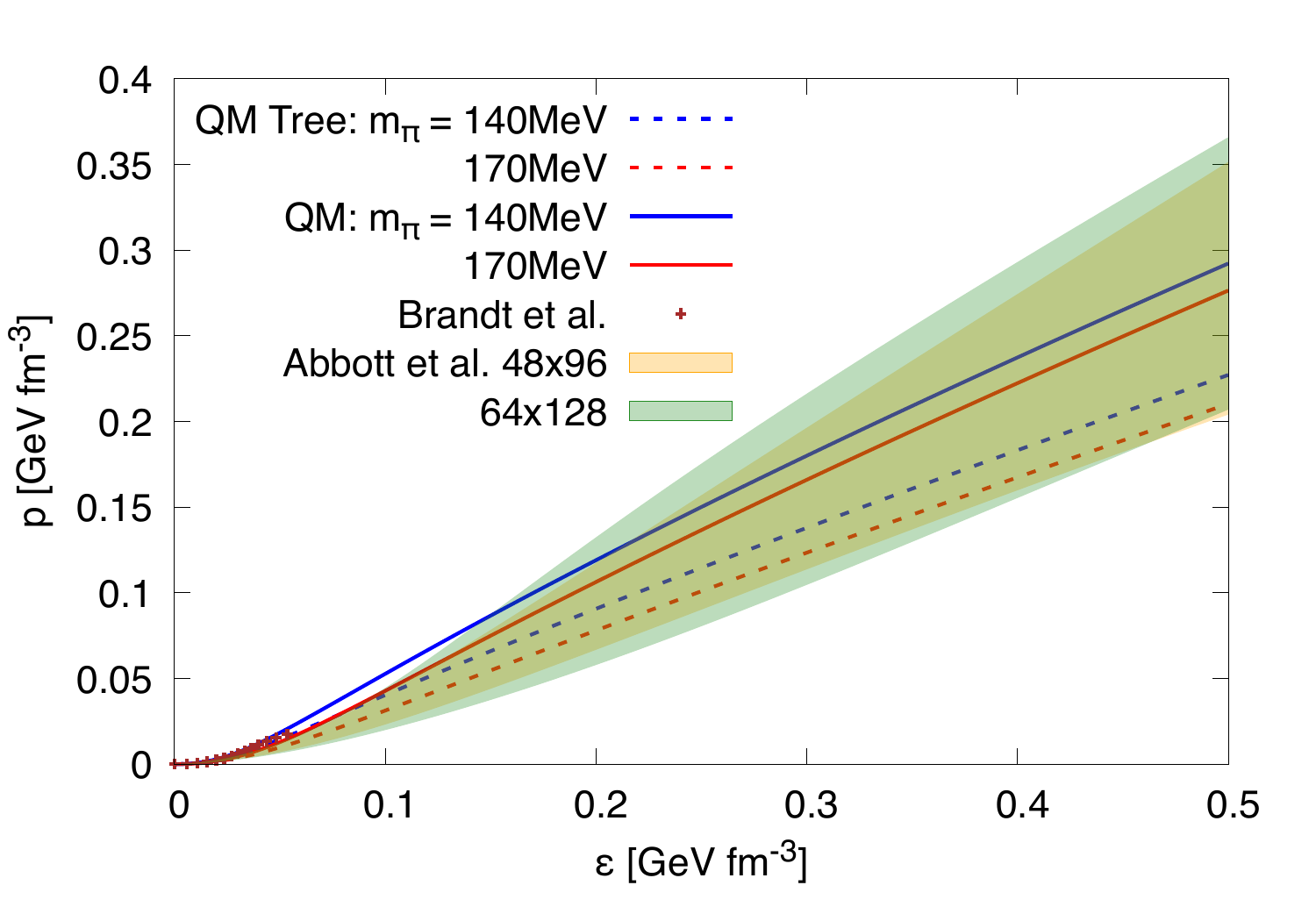}
\includegraphics[width=1.0\linewidth]{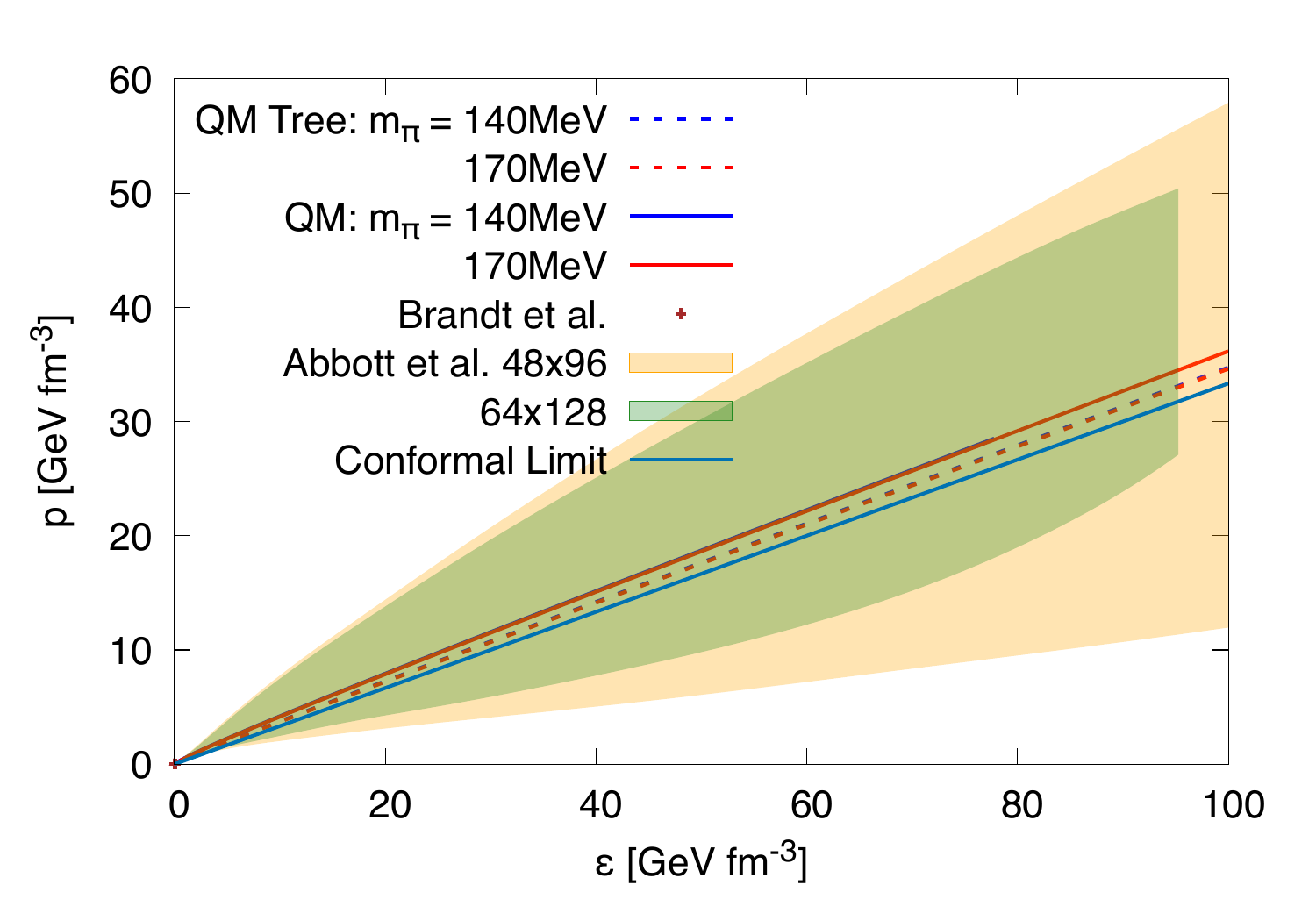}
\vspace{-0.8cm}
\caption{
	Pressure vs energy density from the tree and one-loop effective potential,
	for the low density(upper panel) and global (lower panel) behaviors. 
	}
\label{fig:p_vs_e}
\end{figure}

Shown in Fig.~\ref{fig:p_vs_e} is the pressure as a function of energy density.  
For the density range to $\sim 10n_0$,  
the pressure with quark loops is larger than that that in the tree level by 10-20\%. 
This means that the quark substructure effects enhance the pressure from the purely hadronic one. 
On the other hand, toward high density the difference in $P$ vs $\varepsilon$ becomes much smaller than in $n_I$ vs $\mu_I$.
Such degeneracy is reached when $P$ enters the $\mu_I^4$ scaling regime;
for whatever coefficients of the $\mu_I^4$ term, $P\simeq \varepsilon/3$ is achieved when $\mu_I^4$ terms dominate.

We also note that the pressure is reduced for larger $m_\sigma$ and $f_\pi$, 
as shown in Fig.~\ref{fig:p_vs_e_fpi_dep}.
In other words, with stronger chiral symmetry breaking in vacuum (which increases both $m_\sigma$ and $f_\pi$),
the high density EOS after the chiral restoration becomes softer.
This point is examined in Sec.\ref{sec:ChRes}.

%
\begin{figure}[htbp]
\vspace{-0.cm}
\centering
\includegraphics[width=1.0\linewidth]{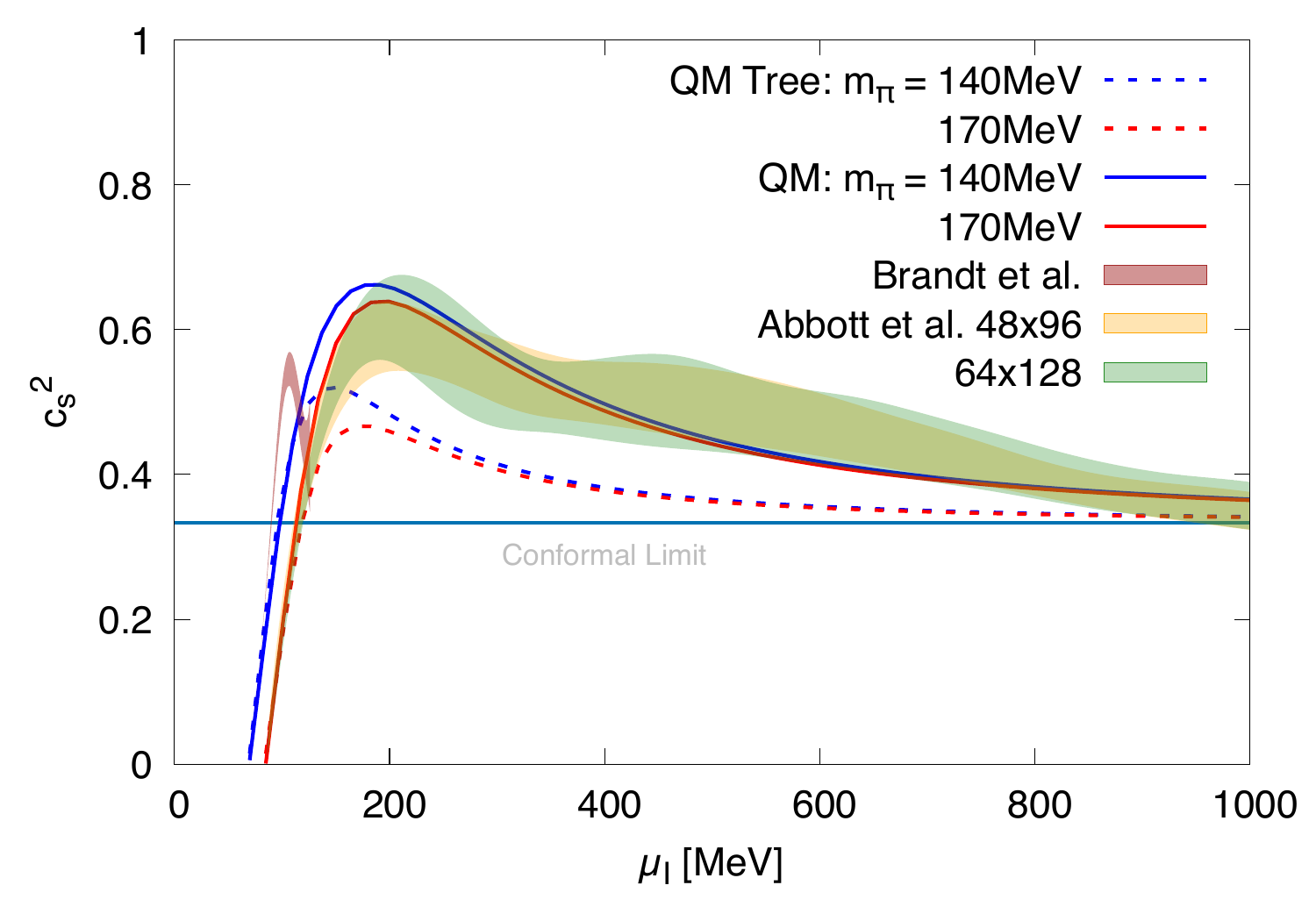}
\includegraphics[width=1.0\linewidth]{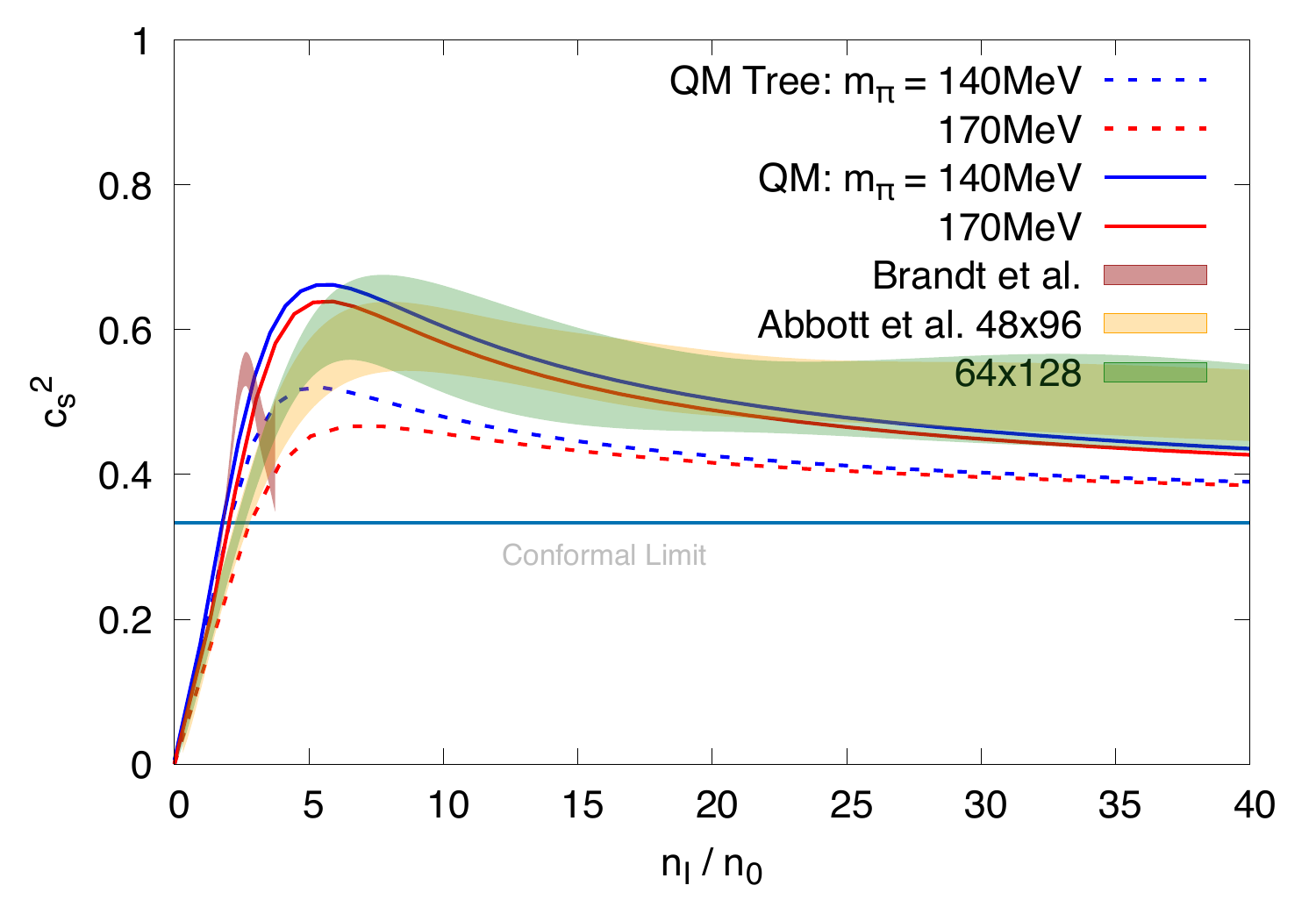}
\vspace{-0.8cm}
\caption{
Squared sound velocity $c_s^2$ vs isospin chemical potential 
(upper panel) and vs isospin density (lower panel).
}
\label{fig:cs2}
\end{figure}
To further examine the variation of stiffness, we now turn to the behavior of the sound velocity.
Figure \ref{fig:cs2} shows the $c_s^2$ as a function of isospin chemical potential
and also as a function of isospin density. 
The $c_s^2$ increases rapidly at low density, makes a peak, and slowly relaxes to the conformal limit $1/3$ from above.
This qualitative feature seems robust and is consistent 
 with the lattice results in Refs.~\cite{Brandt:2022hwy} and \cite{Abbott:2023coj}.
However, the quantitative agreement beyond the BEC regime depends on the lattice results.
The location of the $c_s^2$ peak is near $\mu_I \simeq 1.2 m_\pi$ for $n_I \simeq 5n_0$
 in our calculations for the reasonable range of our parameter set for $m_\sigma$ and $f_\pi$.
The lattice results in Ref.~\cite{Brandt:2022hwy} indicate the peak at $\mu_I \simeq 0.8m_\pi$ or $n_I \sim 0.5 n_0$, lower than our model results.
Meanwhile our results agree better with the results of Ref.~\cite{Abbott:2023coj}, 
although our sound velocity peak is located at density slightly lower than found in the lattice simulations, 6-7$n_0$.
We are not sure about the origin of the discrepancy between results of Refs.~\cite{Brandt:2022hwy} and \cite{Abbott:2023coj} as they seem to contain different systematic errors.
But after performing several parametric studies as given in Sec.~\ref{sec:discussion},
we could not find any {\it qualitative} mechanisms to reconcile $\Delta \simeq \lqcd$ with the quick reduction of $c_s^2$ after making the peak as seen in Ref.~\cite{Brandt:2022hwy}. 
Here we assume $\Delta \simeq \lqcd$ based on the lattice result for the melting temperature of pion condensates, $T_c^{\rm lat} \simeq 170$ MeV,
which seems more or less constant to $\sim n_0$ or even higher densities, see discussions around Eq.\eqref{eq:Tc_lat}.
For this reason, in the beyond-BEC regime, the results of Ref.~\cite{Abbott:2023coj} seem more natural to us than those of Ref.~\cite{Brandt:2022hwy}
whose simulations are more optimized for the low density region.

We note that the $c_s^2$ at tree level also shows the peak in the crossover region and then the convergence to the conformal limit at high density.
As we noted in the Introduction, these behaviors can be acheived
by $P(\mu_I) \sim c_2 \mu_I^2 \Lambda^2 + c_4 \mu_I^4$ with $\Lambda$ being some scale.
Purely hadronic models may achieve this condition, 
but this by itself does not mean that the EOS is described correctly, as we have mentioned in discussion of $P$ vs $\varepsilon$ and $n_I$ vs $\mu_I$.
In our standpoint, the tree level results, which crucially depend on the scaling $\Delta_{\rm tree} \sim \mu_I$, 
becomes potentially misleading at high density.

With the above qualifications in mind,
in the next section 
we look into more details of our model regarding it as a model of composite particles.

\subsection{Occupation Probability}
\label{sec:occupation}

At low density the effective degrees of freedom are pions and their internal structure may be ignored.
At higher density, the inter particle distance becomes shorter and the quark substructure of pions becomes important.

To estimate where the quark substructure becomes important, we refer to the pion charge radius. 
It can be extracted from the vector form factor. 
The experimental determination based on the $\pi e$ scattering and the $e^+ e^- \rightarrow \pi^+ \pi^-$ process \cite{Ananthanarayan:2017efc} yield the estimate
$\la r^2 \ra_V = 0.434(5)\, {\rm fm}^2$ \cite{PhysRevD.98.030001}, or
\beq
r_\pi^V = \sqrt{ \la r^2 \ra_V } ~\simeq~ 0.66\, {\rm fm} \,,
\eeq
which has been well reproduced by lattice calculations \cite{Koponen:2015tkr,Wang:2020nbf}.	
The typical isospin density where pions overlap is estimated through\footnote{In our definition, 
we calculate $n_I = n_u - n_d$, a factor two larger than the conventional definition.
For pions with the isospin 1 to overlap, our $n_I/2$ should be equated with $1/(4\pi r_\pi^3/3)$,
so the factor two must be inserted.}
\beq
\hspace{-0.5cm}
n_I^{\rm overlap} = 2\big( 4\pi r_\pi^3/3)^{-1} \simeq 2 \times 0.83\, {\rm fm}^{-3} \simeq 10.4 n_0 \,.
\eeq
Figure \ref{fig:n_vs_mu} shows that the isospin chemical potential at $n_I^{\rm overlap} $ is 
$\mu_I \simeq$ 200 MeV $\simeq 1.4m_\pi$.

We note that this overlap density $n_I^{\rm overlap} \sim 10n_0 $ is substantially larger than 
the density 2-3$n_0$ where the tree and one-loop results begin to differ substantially,
and the density $\sim 5n_0$ where $c_s^2$ develops a peak.
This would indicate that the quark substructure of hadrons become important before hadrons overlap.
In this respect, there should be a more suitable measure to characterize the location of sound velocity peak. 
One of possible explanations is the {\it quark saturation} \cite{Kojo:2021ugu,Kojo:2021hqh,Kojo:2022psi}.
As density increases, quark states at low momentum are inevitably occupied
and then a newly added quark must fill a state on top of the already occupied states.
Quark states at large momenta are the source of large pressure.

The quark occupation probability in the pion condensed phase can be computed in the standard Nambu-Gor'kov formalism.
The derivation is reviewed in Appendix \ref{appsec:derivation}.
The occupation probabilities for $u$-, $d$-, $\bar{u}$-, $\bar{d}$-quarks are 
\beq
f(p) = f_{u, \bar{d}} (p) &= {1\over 2} \left(1 + { \mu_I - E_D \over E(\mu_I) }\right) \,, \\
\bar{f} (p) = f_{\bar{u}, d} (p) &= {1\over 2} \left(1 + { \mu_I + E_D \over E(\mu_I) }\right) \,.
\eeq
Roughly speaking, $u$- and $\bar{d}$-quarks occupy states up to $\simeq \mu_I$ 
while $\bar{u}$- and $d$-quarks are almost fully occupied as in the Dirac sea without pion condensates.

\begin{figure}[tbp]
\centering
\vspace{-0.cm}
\includegraphics[width=0.85\linewidth]{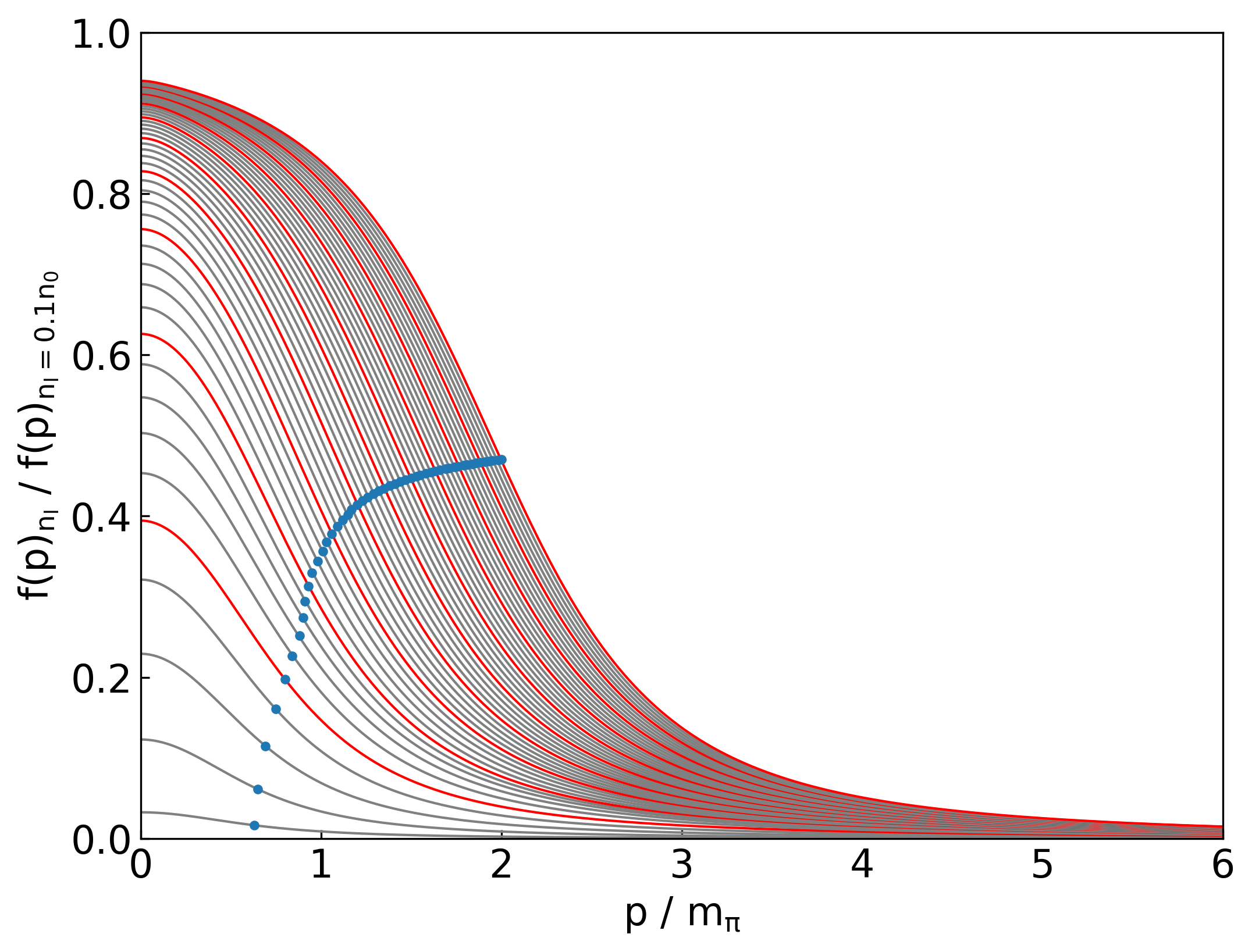}
\includegraphics[width=0.85\linewidth]{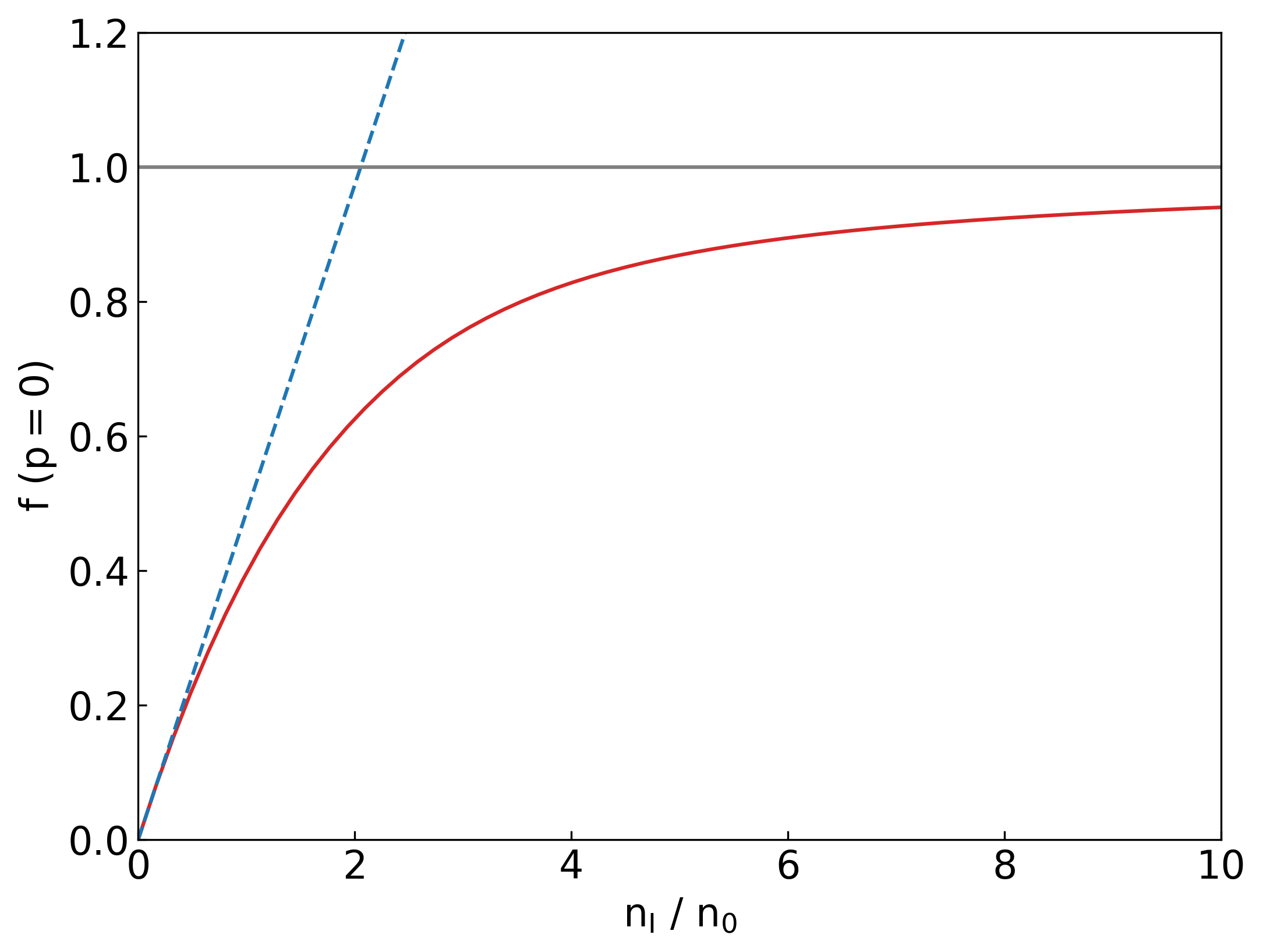}
\caption{
(Upper) The occupation probability of $u$- and $\bar{d}$ quarks which corresponds to the residue of positive energy part of $\langle u\overline{u}\rangle$. 
The densities we plotted are from $0.2$ to $10.0n_0$ in $0.2n_0$ increments for gray curves and $1.0n_0$ increments for red curves.
The blue dots are the locations of the surface of the distribution where $f(p)$ has the half maximum.
(Lower) The evolution of the occupation probability for the $p=0$ state, $f(p=0)$.
}	
\label{fig:occ}
\end{figure}

\begin{figure}[tbp]
\centering
\vspace{-0.2cm}
\includegraphics[width=0.9\linewidth]{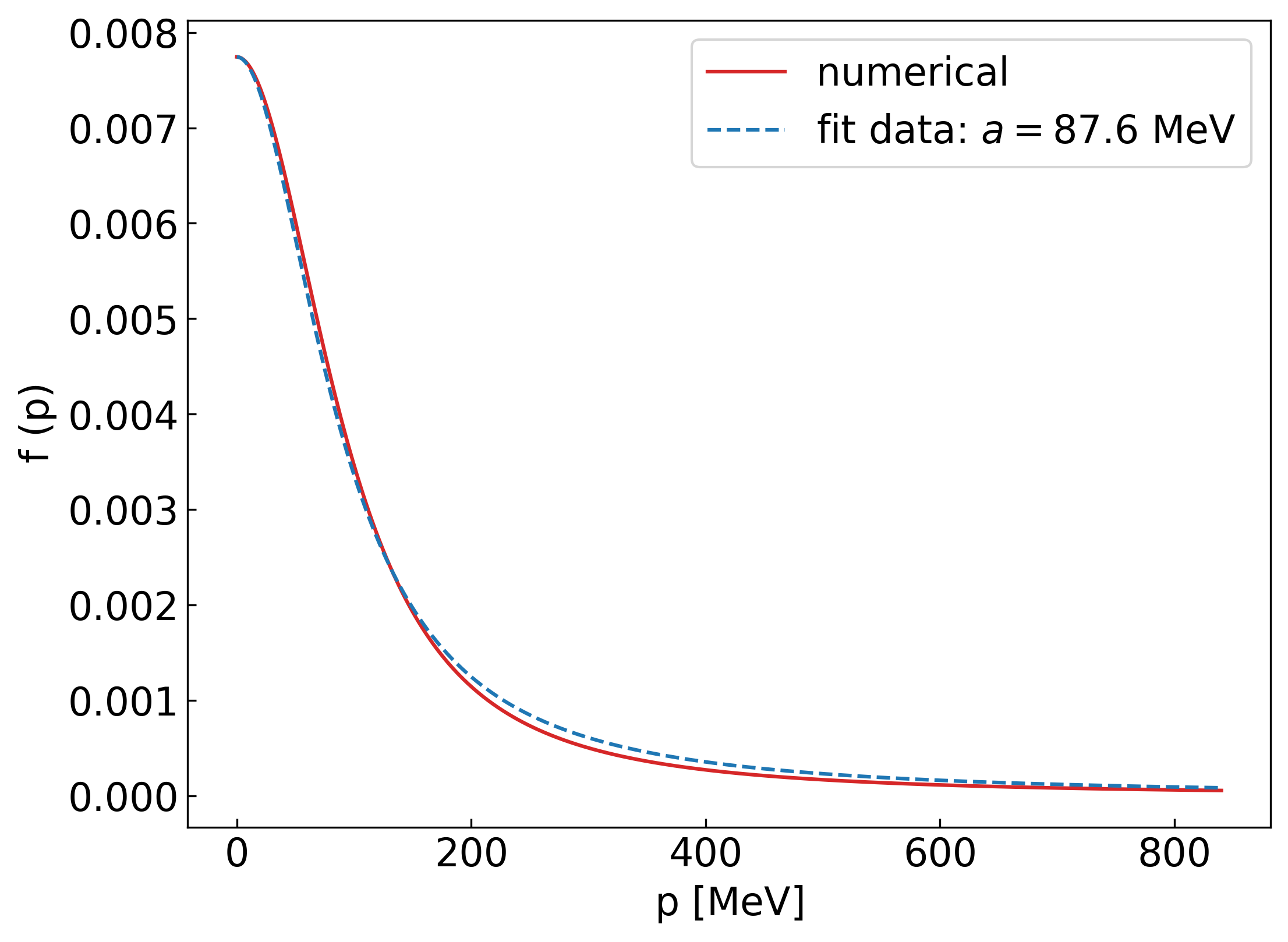}
\caption{
The fit of the dipole function in Eq.\eqref{eq:dipole} to the $\varphi_\pi^{\rm vac}$ calculated in the quark-meson model.
}	
\label{fig:occ}
\end{figure}

Shown in Fig.~\ref{fig:occ} is the occupation probability $f (p) $ at various densities as a function of quark momenta $p$.
The densities we plotted are from $0$ to $10.0n_0$ in $0.2n_0$ increments for gray curves and $1.0n_0$ increments for red curves.
The blue dots, where $f(p)$ takes the half value of $f(p=0)$, are the measure of typical momentum at the Fermi surface.

For later convenience we define the quark distribution in a single pion as
$\varphi_\pi^{\rm vac}(p)  \equiv \lim_{n_I\rightarrow 0} f(p)/n_I $.
It turns out that $\varphi_\pi^{\rm vac} (p)$ is approximated well 
by a simple monopole Ansatz or Breit-Wigner form
\begin{align}
	\varphi_\pi(p) \sim \frac{1}{\, 1 + p^2/a^2 \,} 
	\label{eq:dipole}
\end{align}
with $a \simeq 87.6$ MeV.
This suggests that, in our quark-meson model, 
the pion wavefunction in the coordinate space has the exponentially-decaying form.

\begin{figure}[tbp]
\centering
\vspace{-0.3cm}
\includegraphics[width=1.0\linewidth]{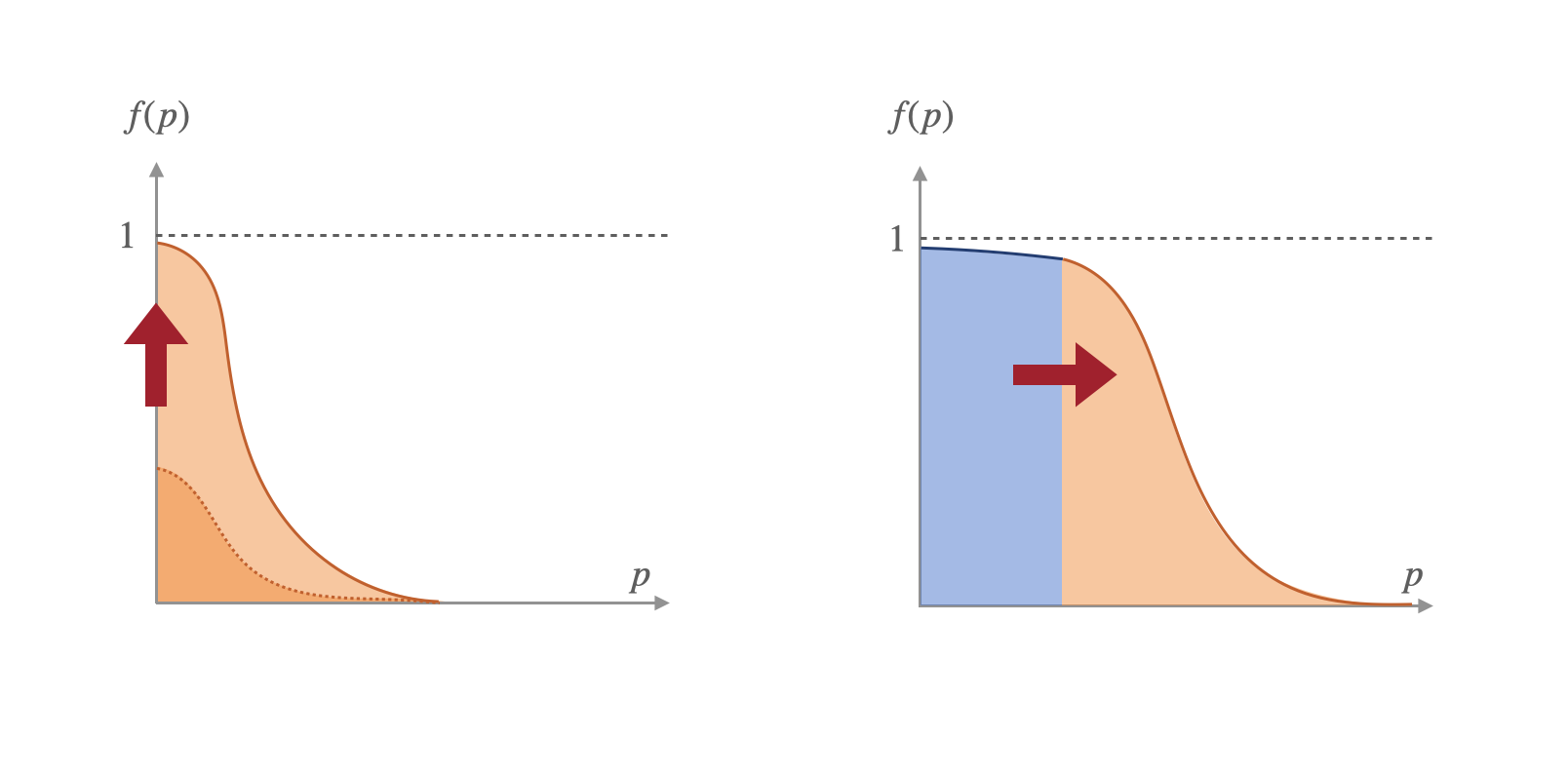} 
\vspace{-1.0cm}
\caption{
	Schematic figures for the evolution of the occupation probability.
		{\bf Left}: The ``vertical'' evolution; 
		{\bf Right}: The ``horizontal" evolution.
		}
\label{fig:occ_cartoon}
\end{figure}

To relate the evolution of $f(p)$ to the stiffening of matter,
it is useful to decompose the evolution of $f(p)$ into two components. 
The first is the ``vertical evolution'' 
in which $f(p)$ just increases its magnitude as $f(p) \simeq n_I \varphi_\pi^{\rm vac} (p)$ 
(Fig.\ref{fig:occ_cartoon}, {\bf Left});
this corresponds to the regime where pions do not interact and quarks inside of pions are largely unaffected.
In this regime,  $\varepsilon/n_I $ is close to a constant, and therefore the pressure, $P = n_I^2 \partial (\varepsilon/n_I)/\partial n_I$, is very small.
While quarks can always contribute to the energy density through the masses of pions,
they do not directly contribute to the pressure. The sound velocity is small in this regime.
The second component is the ``horizontal evolution'' in which
the $f(p)$ increases in the high energy components  (Fig.\ref{fig:occ_cartoon}, {\bf Right}).
This is driven by both interactions and the Pauli blocking effects.
Here, $\varepsilon/n_I$ increases as in usual quark matter and the pressure can be large.
In reality with interactions, the evolution of $f(p)$ is the mixture of these two components.

In our quark-meson model, 
Fig.\ref{fig:occ} suggests that, 
from $0$ to $\sim 2n_0$, the magnitude of $f (p)$ at $p=0$ grows rapidly from $0$ to $\simeq 0.6$,
but at higher density the distribution $f(p)$ develops toward the horizontal direction.
If we treated pions as if elementary and non-ineteracting particles, the $f(p=0)$ would violate the Pauli principle around $\simeq 2n_0$.
The $c_s^2$ peak is located around $\simeq 5n_0$ where $f (p=0) \simeq 0.9$.
Beyond this density the horizontal evolution dominates over the vertical evolution and
$c_s^2$ relaxes toward $1/3$ as in a relativistic quark gas.

We note that, the quark substructure effects are already significant at 1-2$n_0$
and develops a peak in $c_s^2$ at $\sim 5n_0$, 
at density substantially smaller than 
the naive estimate of the pion overlap, $n_I^{\rm overlap} \sim 10 n_0$.
This suggests that the evolution of the occupation probability can represent two characteristic scales;
one is for the quark saturation, and the other is for the overlap of composite particles.
The distinction of such two scales was emphasized in Ref.~\cite{Fukushima:2020cmk}
which discriminates the mode-by-mode percolation in momentum space from the conventional geometric percolation.

Finally we comment on some difference between nucleonic matter and pionic matter in isospin QCD.
In nuclear matter the evolution of $c_s^2$ is much slower than in isospin QCD,
$c_s^2 \lesssim 0.1$ for $n_B \sim 1$-$2n_0$ \cite{Drischler:2021bup}. 
Nucleons are much heavier than pions and $c_s^2$ is naturally small because of the non-relativistic regime.
Two- and three-nucleon repulsions increase $c_s^2$, but their effects basically enter as powers of $\sim n_B^2$ and $\sim n_B^3$
whose growth are rather slow and $c_s^2$ goes beyond $1/3$ only at $n_B \gtrsim $ 2-3$n_0$.
This aspect differs from pionic matter in isospin QCD
where pions can be relativistic already at $\simeq n_0$ and $c_s^2 \ge 1/3$ is achieved already at $n_I \sim 2n_0$.

\section{Discussion}
\label{sec:discussion}

Here we address several issues not detailed in the previous sections.
First we discuss how the strength of chiral symmetry breaking in vacuum and
its restoration at high density affect EOS.
For the high density domain, we compare our results with pQCD at high density, and conjecture the importance of the power corrections.
Then we discuss the trace anomaly and the positivity conjecture.

\subsection{Chiral symmetry restoration and softening}
\label{sec:ChRes}

In Sec.\ref{sec:eos} we have seen that larger $f_\pi$ and/or $m_\sigma$ lead to softer EOS at high density.
Here we try to explain this softening by focusing on the chiral symmetry breaking in the vacuum and its restoration at high density.
In this context larger $f_\pi$ and $m_\sigma$ mean the stronger chiral symmetry breaking in the QCD vacuum. 

\begin{figure}[tbph]
\vspace{-0.cm}
\centering
\includegraphics[width=0.9\linewidth]{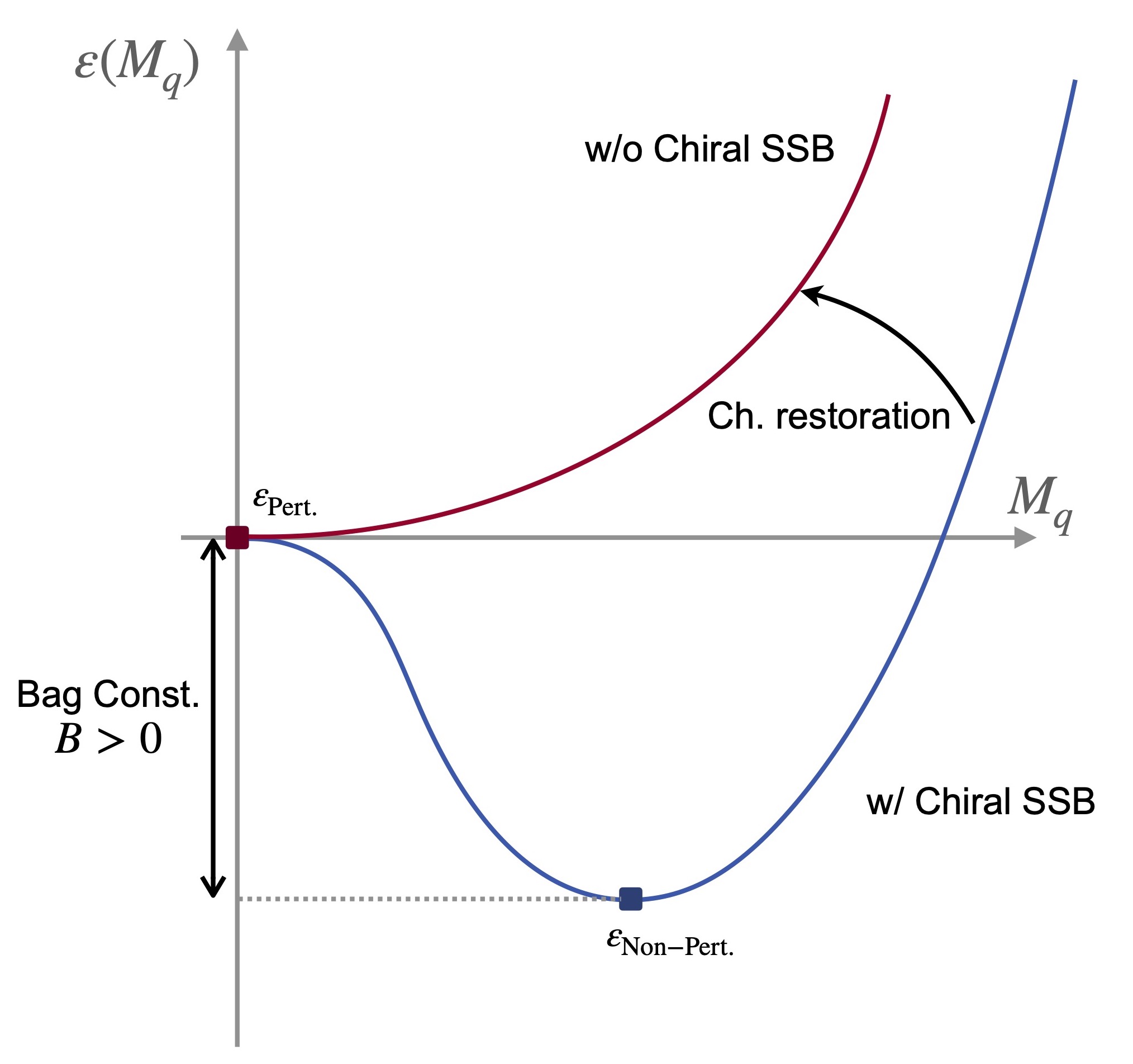}
\caption{  The energy density as a function of the chiral effective mass $M_q$.
		After chiral restoration the minimum energy is realized at $M_q=0$.
		In the broken phase
		the minimum is realized at $M_q \neq 0$ and the energy is smaller than that of $M_q=0$.
		This gap in the zero-point energy density is the bag constant.
	}
\label{fig:bagconst1}
\end{figure}

\begin{figure}[tbph]
\vspace{-0.cm}
\centering
\includegraphics[width=0.9\linewidth]{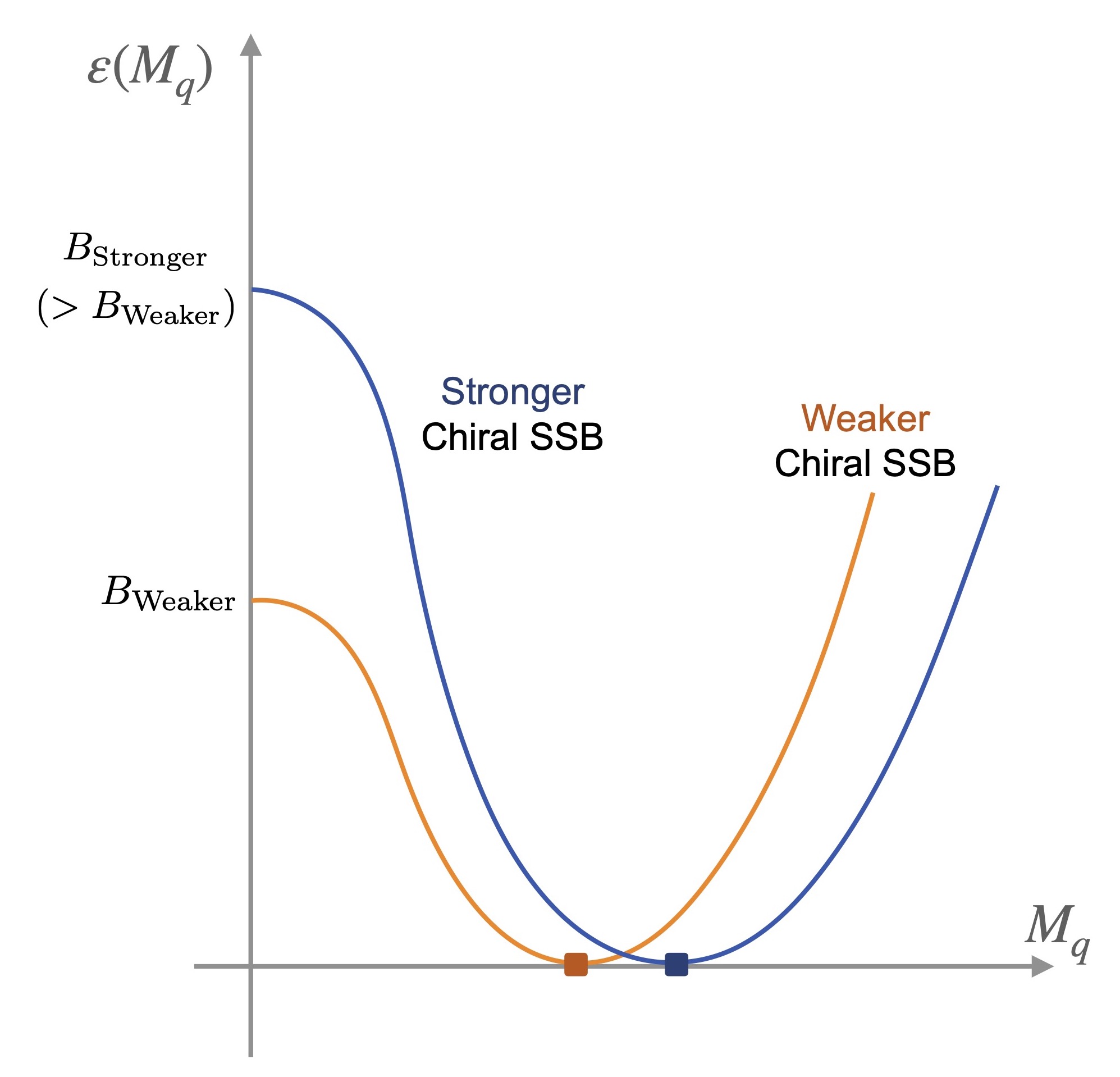}
\caption{ The energy density with different strength of the chiral symmetry breaking.
		The bag constant is larger for the stronger chiral symmetry breaking. 
		}
\label{fig:bagconst2}
\end{figure}

In the vacuum, the energy reduction due to the chiral symmetry breaking is  (Fig.\ref{fig:bagconst1})
\beq
B \equiv  V_{\rm 1-loop} (M_q = 0 ) - V_{\rm 1-loop} (M_q = M_0 ) \,,
\eeq
where the first term is the energy of the trivial vacuum while the second one is the energy of the chiral symmetry broken vacuum. 
This sort of the energy difference is often called the bag constant.
Stronger breaking in the chiral symmetry increases the size of the bag constant (Fig.\ref{fig:bagconst2}).
In our model the bag constants $B(m_\sigma,f_\pi)$ 
in the $m_\pi^4 = (140\, {\rm MeV})^4 \simeq 50.25\, {\rm MeV fm^{-3} }$ unit are given as
\begin{align}
	B(450, 90) = 0.874, ~~~ B(450, 100) = 1.035 , \notag\\
	B(600, 90) = 1.439, ~~~ B(600, 100) = 1.698, 
\end{align}
from which one can see that larger  $f_\pi$ and $m_\sigma$, i.e., stronger chiral symmetry breaking, 
lead to a greater $B$.

A larger bag constant softens EOS at high density. 
To see this, it is useful to recall a bag model with perturbative corrections.
We note that the perturbative expansions are performed around the trivial vacuum.
Since our EOS is normalized to make $P=0$ at $\mu=T=0$ for the non-perturbative vacuum,
the perturbative evaluation of EOS must be corrected by the non-perturbative normalization constant.
Then, the pressure and energy density are
\beq
\hspace{-0.5cm}
P_{\rm pert}^{\rm normalized} = P_{\rm pert} - B, ~~ \varepsilon_{\rm pert}^{\rm normalized} = \varepsilon_{\rm pert} + B.
\eeq
The bag constant associated with the chiral restoration reduces the pressure and increases the energy density, 
resulting in a softer EoS at high density where the chiral symmetry is restored.
Similar conclusions have been obtained in models with and without the $U(1)_A$ anomaly \cite{Gao:2022klm,Minamikawa:2023eky}.

%



%
%

\subsection{Power corrections to pQCD at high density}
\label{sec:compare_pQCD}

Our quark-meson EOS predicts $c_s^2$ approaching $1/3$ from above as density increases.
This contradicts with the pQCD prediction in which $c_s^2$ approaches $1/3$ from below.
A possible origin of such discrepancy would be the power corrections of $\sim \lqcd^2 \mu_I^2$ 
which cannot be derived from perturbative computations.

In Introduction, we schematically showed how power corrections can enhance
the $c_s^2$ in Eqs.(\ref{eq:pressure_intro}) and (\ref{eq:cs2_intro}).
The question is how large power corrections should be to qualitatively change
the perturbative behaviors of $c_s^2$.

For a given flavor $f$, the pQCD EOS up to $O(\alpha_s^2)$ is given as
\cite{Graf:2015pyl} 
(we use the current quark mass, $m_{u,d} \simeq 5$ MeV
and $\mu_I = \mu_u = - \mu_d$ in the present work)
\beq
P_0^f 
\! &=&\! \frac{\Nc}{\, 12\pi^2 \,} \bigg[
|\mu_f| u_f  \bigg( \mu_f^2 - \frac{\, 5 \,}{2} m_q^2 \bigg) 
+ \frac{\, 3 \,}{2} m_q^4 \ln \frac{\, |\mu_f| + u_f \,}{ m_q } \bigg] \,,
\notag
\\
P_1^f 
\! &=& \!
- \frac{\, \alpha_s N_{\rm G} \,}{\, 16 \pi^2 \,} \bigg[
3 \bigg( m_q^2 \ln \frac{\, |\mu_f| + u_f \,}{ m_q }  - |\mu_f| u_f \bigg)^2 
- 2 u_f^4
\notag 
\\
&&\hspace{-0.7cm} +  
 m_q^2 \bigg( 6 \ln \frac{\, \Lambda_{\rm reno} \,}{\, m_q \,} + 4 \bigg) 
	\bigg( |\mu_f| u_f - m_q^2 \ln  \frac{\, |\mu_f| + u_f \,}{ m_q } \bigg) 
\bigg] \,. \notag \\
\eeq
In the isospin symmetric limit, $P^u = P^d$.
Here, $P_0^f$ and $P_1^f$ is the zeroth and first order in $O(\alpha_s) $,
with $u_f = \sqrt{ \mu_f^2 - m_q^2 } $, $N_{\rm G} = \Nc^2-1$,
and $\Lambda_{\rm reno}$ being the renormalization scale.
The running $\alpha_s$ is
\beq
\alpha_s (\Lambda_{\rm reno})
&=& \frac{\, 4 \pi \,}{\, \beta_0 L \,} \bigg[ 1 - 2 \frac{\, \beta_1 \,}{\, \beta_0^2 \,} \frac{\, \ln L \,}{\, L \,} \notag\\
&+& \frac{\, \beta_1^2 \,}{\, \beta_0^4 L^2 \,} \big( \ln^2 L -\ln L - 1 + \frac{\, \beta_2\beta_0 \,}{\, \beta_1^2 \,} \big) \notag\\
&+& \frac{\, \beta_1^3 \,}{\, \beta_0^6 L^3 \,} \big( - \ln^3 L + \frac{\, 5 \,}{\, 2 \,}\ln^2 L + 2\ln L \notag\\
&-& \frac{\, 1 \,}{\, 2 \,} - 3 \frac{\, \beta_2 \beta_0 \,}{\, \beta_1^2 \,} \ln L + \frac{\, \beta_3 \beta_0^2 \,}{\, 2\beta_1^3 \,} \big)
\bigg]
\label{eq:pQCD_alpha_s}
\eeq
with $L = 2 \ln( \Lambda_{\rm reno}/\Lambda_{\msbar})$, $\beta_0 = 11-2\Nf/3$, $\beta_1 = 102-19\Nf/3$,
\beq
\beta_2 \! &=& \! \frac{\, 2857 \,}{\, 2 \,} - \frac{\, 5033 \,}{\, 18 \,}N_f + \frac{\, 325 \,}{\, 54 \,}N_f^2\,, \\
\beta_3 \! &=& \! \bigg( \frac{\, 149753 \,}{\, 6 \,} + 3564\, \xi(3) \bigg) \notag\\
\! && \! - \bigg( \frac{\, 1078361 \,}{\, 162 \,} + \frac{\, 6508 \,}{\, 27 \,}\xi(3) \bigg) N_f \notag\\
\! && \! + \bigg( \frac{\, 50065 \,}{\, 162 \,} + \frac{\, 6472 \,}{\, 81 \,}\xi(3) \bigg) N_f^2 \notag\\
\! && \! + \frac{\, 1093 \,}{\, 729 \,} N_f^3
\eeq
and $\Lambda_{\msbar} \simeq 340\, {\rm MeV}$.
The central value of $\Lambda_{\rm reno}$ is $\Lambda_{\rm reno} = \mu_f$,
and as usual we vary $\Lambda_{\rm reno} $ from $ \mu_f$ to $4\mu_f$.

\begin{figure}[tbp]
\vspace{-0.5cm}
\centering
\includegraphics[width=1.05\linewidth]{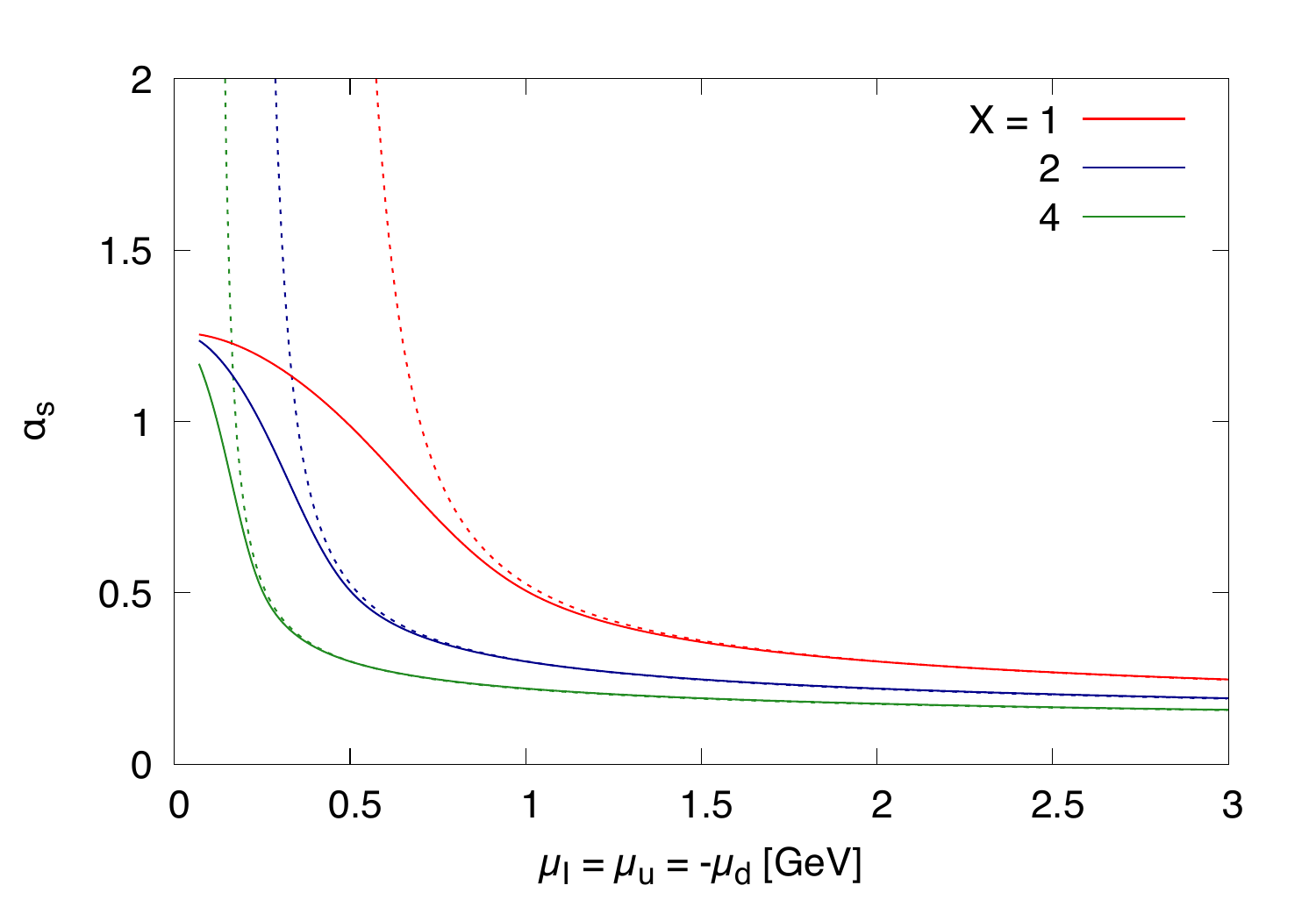}
\caption{ 
Running coupling from pQCD (dashed) and from the freezing coupling (solid).
For the renormalization scale $\Lambda_{\rm reno} = X \mu_I$,
we examine $X= 1, 2$, and $4$.
		}
\label{fig:alphas_pQCD}
\end{figure}

In addition to the perturbative running coupling which becomes unphysical toward the Landau pole,
we also examine the case with the freezing coupling in the low energy limit.
We divide the domain into three
\beq
\alpha_s(Q^2)
&=& \alpha^{\rm low}_s(Q^2) \theta(t_{\rm low}-Q^2)   \notag \\
&+& \alpha^{\rm mid}_s(Q^2) \theta(t_{\rm high}-Q^2) \theta(Q^2-t_{\rm low})  \notag \\
&+& \alpha^{\rm high}_s(Q^2) \theta(Q^2 -t_{\rm high}) \,.
\eeq
where $t_{\rm low}^{1/2} = 0.3\, {\rm GeV}$ and $t_{\rm high}^{1/2} = 1.1$ GeV.
For the low energy limit we use the form suggested by Deur et al \cite{Deur:2014qfa,Deur:2016cxb} 
\beq
\alpha_s^{\rm low} 
= \alpha_s^{\rm low} (0) \rme^{-Q^2/ 4\kappa^2} \,,
\eeq 
with $\alpha_s^{\rm low} (0) \simeq 1.22$ and $\kappa \simeq 0.51$.
For the high density we use the perturbative expression (\ref{eq:pQCD_alpha_s}),
and for the intermediate region we use the interpolant
\beq
\alpha_s^{\rm mid} (Q^2) = \sum_{m=0}^5 c_m \mu_I^m \,,
\eeq
where $c_n$'s are fixed by demanding the matching
\beq
\frac{\, \partial^n \alpha_s^{\rm low/high} }{ (\partial Q^2)^n } \bigg|_{ Q^2 = t_{\rm low/high} }
= \frac{\, \partial^n \alpha_s^{\rm mid} }{ (\partial Q^2)^n } \bigg|_{ Q^2 = t_{\rm low/high} } \,,
\eeq
for $n=0,1,2$. The six boundary conditions fix the six $c_n$ uniquely. 
Unlike in Refs.~\cite{Deur:2014qfa,Deur:2016cxb} which needed only the continuity up to the first derivative, 
in this work we use the interpolant not to generate any discontinuities up to the second derivative,
since we compute $c_s^2$.

\begin{figure}[tbp]
\vspace{-0.5cm}
\centering
\includegraphics[width=1.05\linewidth]{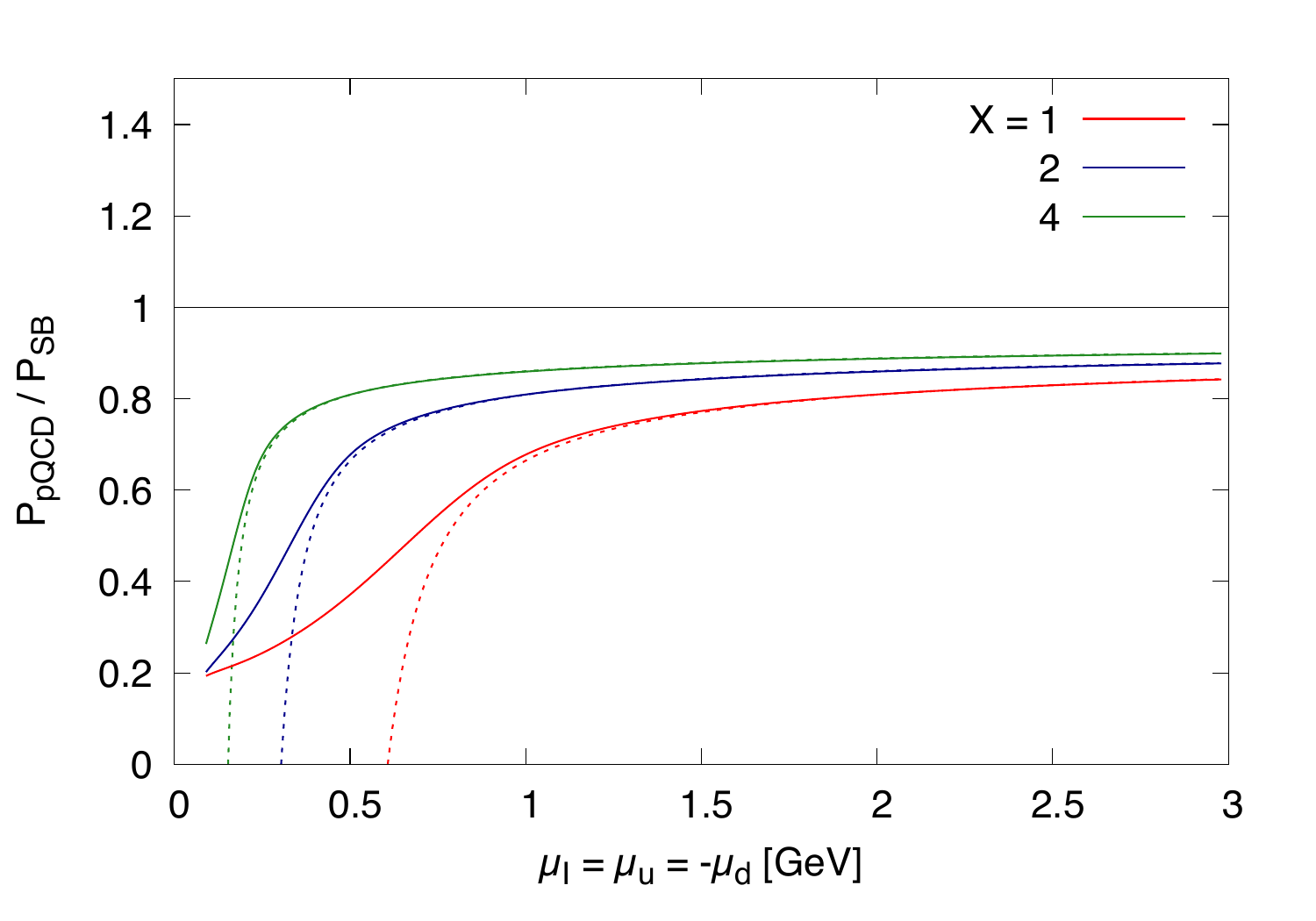}
\vspace{-0.8cm}
\caption{ 
Perturbative pressure with perturbative running with the Landau pole and infrared freezing coupling.
The pressure is normalized by the pressure in the Stefan-Boltzmann limit.
Notations for the solid and dashed lines are the same as Fig.~\ref{fig:alphas_pQCD}.
}
\label{fig:pQCD_P_vs_mu_D0}
\end{figure}

We set $Q^2=\Lambda_{\rm reno}^2$ and plot $\alpha_s (\Lambda_{\rm reno})$ in Fig.\ref{fig:alphas_pQCD} together with the pQCD running coupling.
With the IR freezing coupling the artificial reduction of pQCD pressure is tempered 
and the pressure remains positive toward the low density region (Fig.\ref{fig:pQCD_P_vs_mu_D0}).

Now we add power corrections which are parametrized in terms of gaps in the pion condensed phase.
The phase space factor $\sim 4\pi p_F^2 \Delta$ times the gap $\Delta$, divided by a factor $(2\pi)^3$, yields 
the naive estimate
\beq
P_{\rm cond}
= C
\frac{\, \mu_I^2 \Delta^2 \,}{\, \pi^2 \,} \,,
\eeq
where $C$ is a constant of $O(1)$.
For our quark meson model $C \simeq \Nc/2$, see Eq.~\eqref{eq:power_from_gap}.

In Son's estimate \cite{Son:1998uk}, 
based on the color-magnetic long range forces,
the gap is evaluated as
\beq
\Delta_{\rm color-mag} = b |\mu_I | g_s^{-5} \rme^{- 3\pi^2/2 g_s }
\eeq
with $b\sim 10^4$
and $g_s = g_s (|\mu_I|)$ being the running coupling constant.
The gap can be several hundreds MeV.
Meanwhile, in our quark-meson model, we have found $\Delta \simeq 300$ MeV.
It is interesting to note that such $\Delta$ seems to satisfy the BCS relation
between the gap and the critical temperature,
\beq
T_c^{\rm BCS} \simeq 0.57\, \Delta_{\rm BCS} \,,
\label{eq:Tc_lat}
\eeq
which implies $T_c \simeq 171\, {\rm MeV}$, 
in good agreement with the lattice result $T_c^{\rm lattice} \simeq 160$-$170\, {\rm MeV}$
for the interval $\mu_I \simeq 100$-$300\, {\rm MeV}$.

\begin{figure}[tbp]
\vspace{-0.5cm}
\centering
\includegraphics[width=1.0\linewidth]{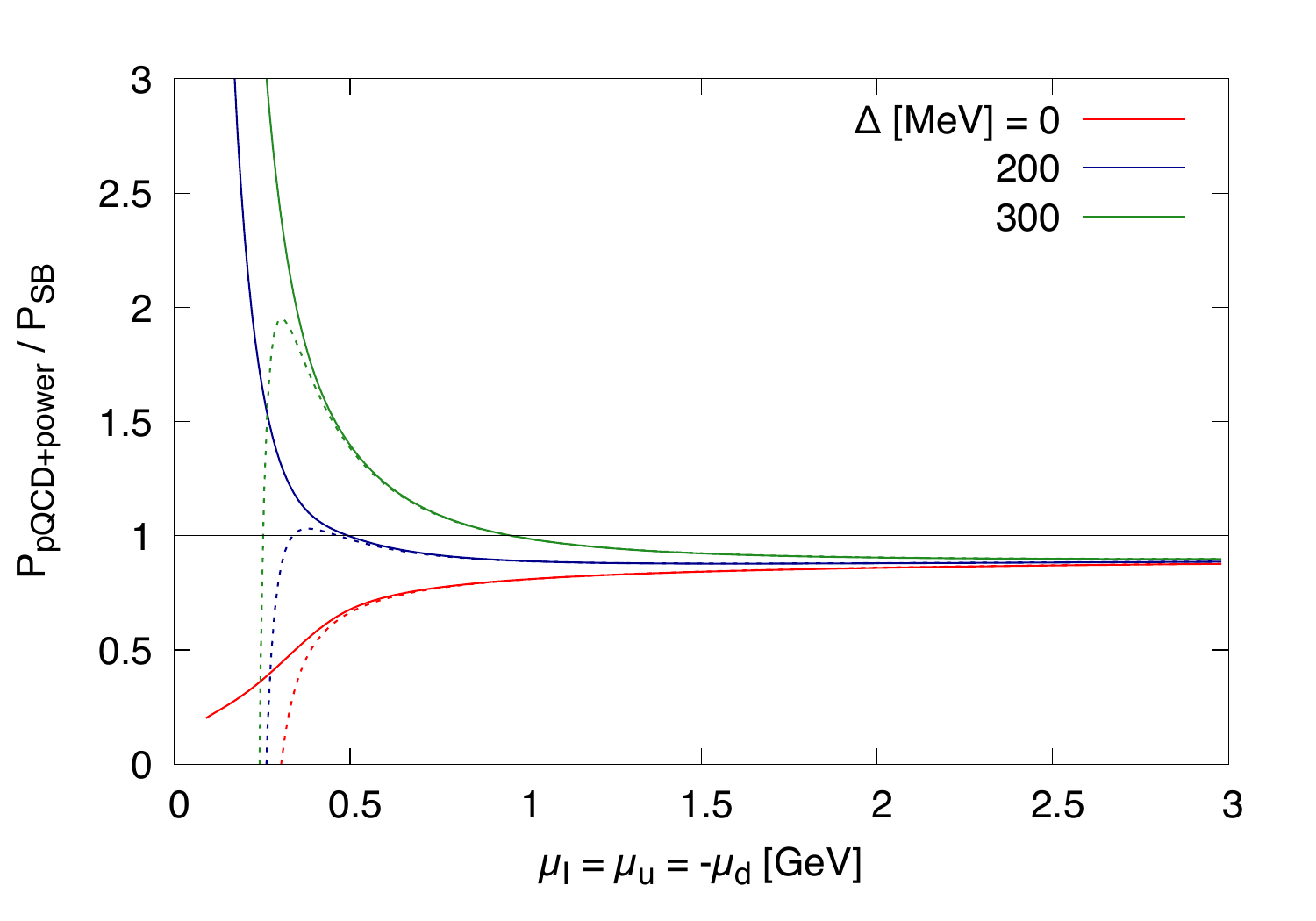} 
\vspace{-0.8cm}
\caption{ 
Perturbative pressure plus power corrections divided by the pressure in the Stefan-Boltzmann limit.
The $\Delta =0, 200$, and $300$ MeV.
The solid and dashed lines represent the freezing coupling and perturbative running with the Landau pole, respectively.
The $X$ for the $\Lambda_{\rm reno}$ is fixed to $X=2$.
		}
\label{fig:pQCD_P_vs_mu_X1}
\end{figure}
\begin{figure}[tbp]
\vspace{-0.5cm}
\centering
\includegraphics[width=1.0\linewidth]{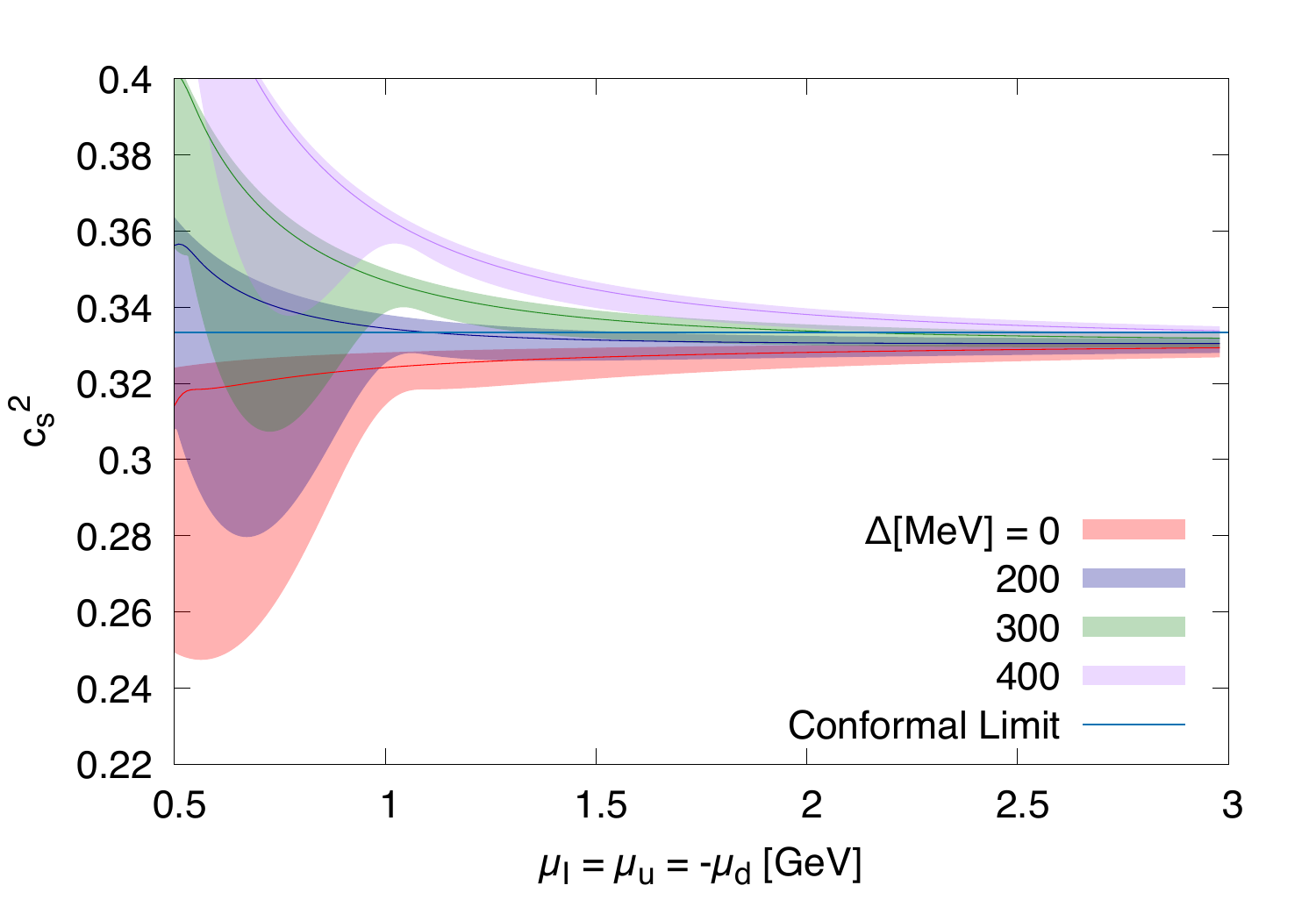} 
\vspace{-0.8cm}
\caption{ 
Squared sound velocity $c_s^2$ for pQCD+power corrections. 
We use the freezing coupling and vary $X$ from \blueflag{$1$ to $4$} to make bands.
The solid lines in the bands are the \blueflag{$X=2$} case.
}
\label{fig:cs2_vs_n_pQCD}
\end{figure}

We set $C=1$ and examine pressure (Fig.~\ref{fig:pQCD_P_vs_mu_X1}) and $c_s^2$ (Fig.~\ref{fig:cs2_vs_n_pQCD}) for $\Delta =0, 200$, and $300$ MeV
for pQCD running coupling (dashed) and freezing coupling (solid).
For $\Delta \simeq 200$ MeV, the power corrections are large enough for $c_s^2$ to approach the conformal limit from above
around $\sim 40 n_0$.
Meanwhile, at $\mu_I \sim 1$ GeV or $n_I \sim 40n_0$, 
parametrically the power corrections in EOS are corrections of the order
\beq
\bigg( \frac{\, \Delta \,}{\, \mu_I \,} \bigg)^2 
= 0.09 \bigg( \frac{\, \Delta \,}{\, 300\, {\rm MeV} \,} \bigg)^2 \bigg( \frac{\, 1\,{\rm GeV} \,}{\, \mu_I \,} \bigg)^2 \,,
\eeq
thus $\sim 10\%$ corrections.
It is remarkable that even such small corrections can change the qualitative behaviors of $c_s^2$
in the domain where pQCD seems applicable.

\subsection{Trace Anomaly}
\label{sec:trace_anomaly}

Recently there has been growing interest in the trace anomaly 
in the context of mechanical properties in hadrons \cite{Fukushima:2020cmk,Polyakov:2002yz,Fujita:2022jus,Sakai:2022zdc} and in neutron stars \cite{Ma:2019ery,Fujimoto:2022ohj,Marczenko:2022jhl}.
The latter is essentially the relation between $P$ vs $\varepsilon$ and is more fundamental than $c_s^2$
which includes only the information of $\rmd P/\rmd \varepsilon$, not the overall magnitude of $P$.
In particular, Ref.~\cite{Fujimoto:2022ohj} conjectured the trace anomaly to be positive.
Below we quickly mention the trace anomaly in dense matter and examine the positivity conjecture 
by considering several non-perturbative effects.

The trace anomaly measures the breaking of the scale invariance and is given by the expectation values of the operator
\beq
\partial_\mu J^\mu_D
= T^{\mu}_{\mu} 
= - \frac{\, \beta(g_s) \,}{2g_s} G^{\mu \nu} G_{\mu \nu} + \sum_f m_f ( 1+ \gamma_m ) \bar{q}_f q_f \,, \notag \\
\eeq
where $J^\mu_D$ is the dilatation current,
$\beta < 0 $ the QCD beta function, and $\gamma_m$ the anomalous dimension of the quark mass. 

For a hadronic state $| K \ra$ with the momentum $K$, the energy momentum tensor gives
\beq
\la K | T^{\mu \nu} (x) | K \ra = K^\mu K^\nu/m_H \,,
\eeq
where the RHS is $x$-independent\footnote{ The state $|K\ra$ with definite momenta is a plane wave,
meaning that the hadron can exist anywhere with the probability $1/V_{\rm space}$.
} and does not contain $g^{\mu \nu}$.
The overall $1/m_H$ factor is fixed by 
the condition at the rest frame, $K_R^\mu = (m_H , {\bf  0} )$,
\beq
\frac{\, \la K_R | H | K_R \ra \,}{\, \la K_R | K_R \ra \,}  
= \frac{\, \la K_R | \int_{\vx} T^{00} (x) | K_R \ra \,}{\, \la K_R| K_R\ra \,}
= m_H \,,
\eeq
where we divide by $\la K_R |K_R \ra$ to cancel the volume factor in the numerator.
Thus, for a hadron at rest frame, we find
\beq
\la K_R | T^{\mu}_{\mu} (x) | K_R \ra = \la K_R | T^{00} (x) | K_R\ra = m_H\,, 
\eeq
with vanishing spatial components, $ \la K_R | T^{ii} (x) | K_R\ra =0$.
The trace anomaly is positive for a single hadron.

It is interesting to extend the above arguments to a many-body system.
Unlike the previous single particle case,
not all particles stay at $\vK=0$.
For instance an ideal Fermi gas leads to
\beq
 \la K_1,\cdots | T^{ii} | K_1, \cdots \ra
\sim \int_{\vK} \frac{\, \vK^2 \,}{\, m_H \,} \,.
\eeq
After lowering one index, we get $\la T^i_i \ra < 0 $.
Thus, the trace anomaly in a many-body system can be negative in principle.
In thermodynamic systems, $\la T^\mu_\mu \ra = \varepsilon - 3P$;
the negative trace anomaly means very large pressure, i.e., stiff EOS.

The trace anomaly characterizes the deviation from the relativistic or conformal limit
as expected at very high density.
We first examine the impact of the normalization in EOS.
In the case of a bag model, we have
\beq
\hspace{-0.5cm}
\la T^\mu_\mu \ra_{\rm bag} 
= \big( \varepsilon  - 3P \big)_{\rm pert}^{\rm normalized}
= \big( \varepsilon  - 3P \big)_{\rm pert} + 4B \,.
\eeq
Changes from the non-perturbative to perturbative vacua enhances the trace anomaly,
supporting the positivity conjecture.
Next we consider the impact of power corrections using EOS similar to Eq.(\ref{eq:pressure_intro}),
\beq
P_{\rm with\,powers} = a_0 \mu_I^4 + a_2 \mu_I^2 \,,
\eeq
but now we include the running of coefficients $a_0$ and $a_2$ associated with $\alpha_s (\mu_I) $.
The energy density can be computed as
\beq
\varepsilon
= 3 a_0 \mu_I^4 + 2 a_2 \mu_I^2 + \frac{\, \partial a_0 \,}{\, \partial \ln \mu_I \,} \mu_I^4 
+ \frac{\, \partial a_2 \,}{\, \partial \ln \mu_I \,} \mu_I^2 \,,
\eeq
and the trace anomaly is
\beq
\hspace{-0.5cm}
\la T^\mu_\mu \ra_{\rm with\,powers}
= - 2 a_2 \mu_I^2 + \frac{\, \partial a_0 \,}{\, \partial \ln \mu_I \,} \mu_I^4 
+ \frac{\, \partial a_2 \,}{\, \partial \ln \mu_I \,} \mu_I^2 \,.
\eeq
The running of $\alpha_s$ favors the positive trace anomaly, 
while the attractive power corrections ($a_2 >0$) favor the negative trace anomaly.

\begin{figure}[tbp]
\vspace{-0.8cm}
\centering
\includegraphics[width=1.05\linewidth]{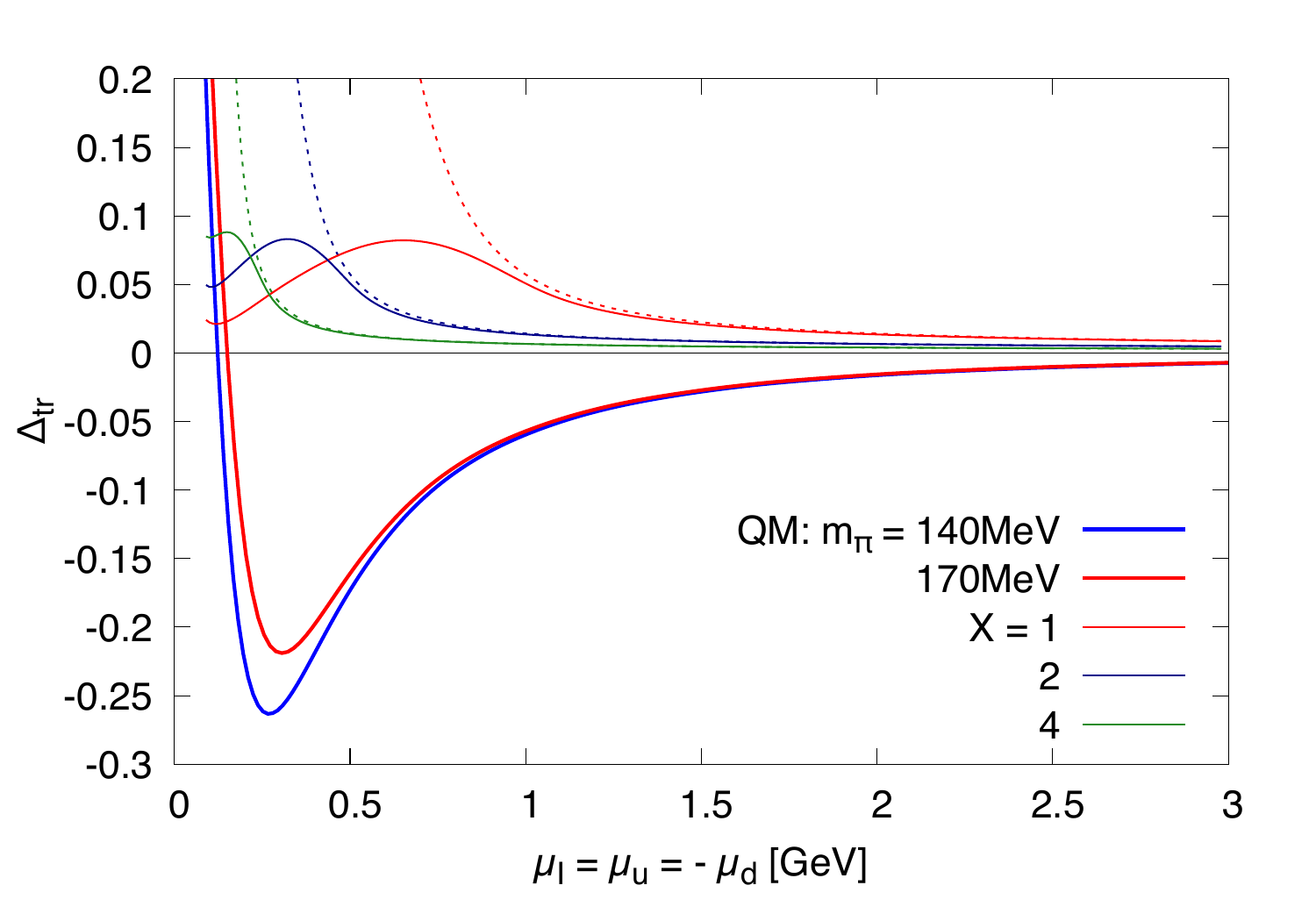}
\vspace{-0.8cm}
\caption{ 
Trace anomaly $\Delta_{\rm tr} = 1/3 - P/\varepsilon$ as functions of $n_I$ for our quark-meson model ($\Nf=2$ QM) and pQCD with the perturbative running coupling (dashed) and IR freezing coupling (solid).
We vary $X$ from 1 to 4.
The trace anomaly is all positive in pQCD but negative for the quark meson model. 
}
\label{fig:pQCD_tr_D0}
\end{figure}

\begin{figure}[tbp]
\vspace{-0.5cm}
\centering
\includegraphics[width=1.0\linewidth]{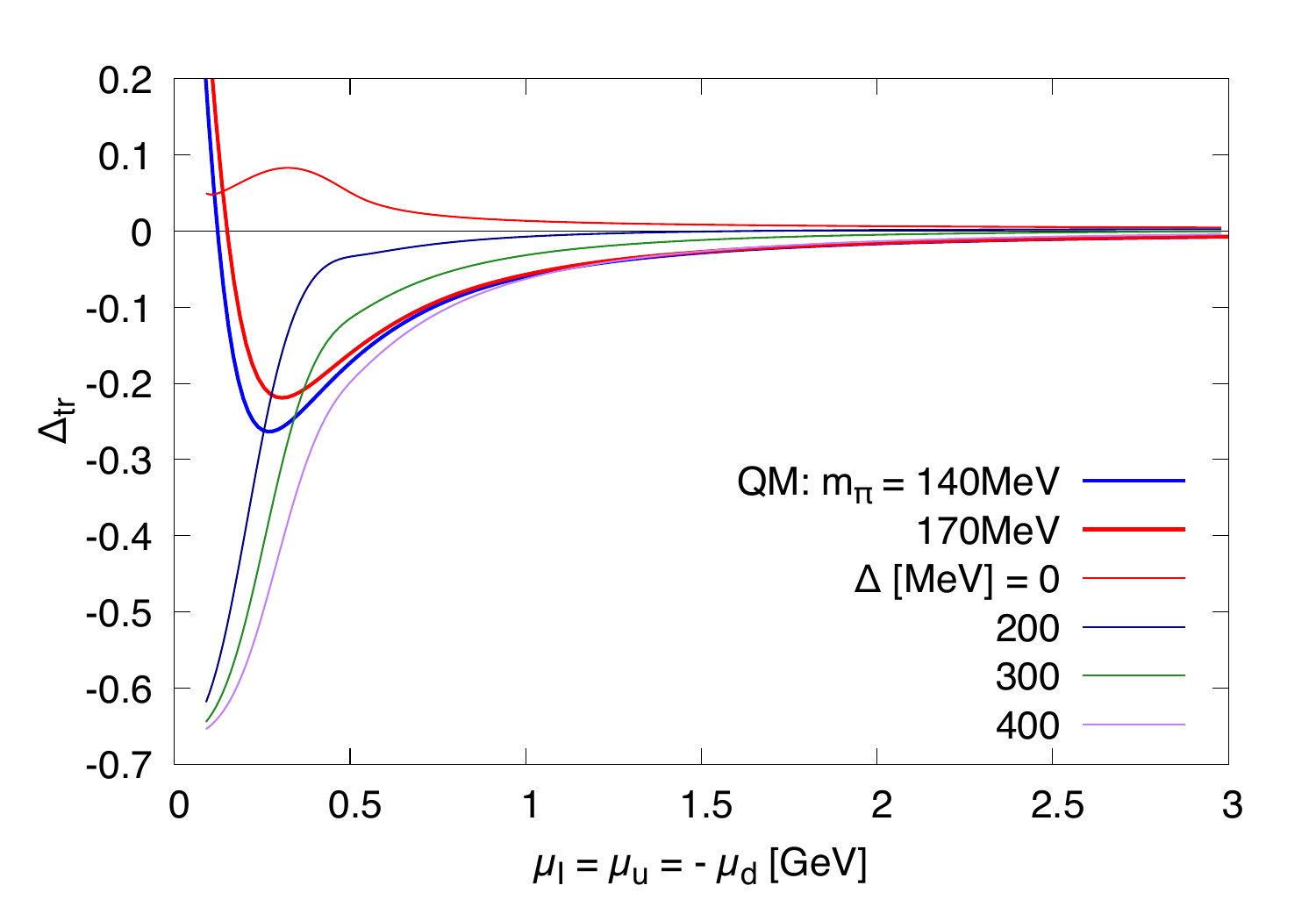}
\vspace{-0.8cm}
\caption{ 
Trace anomaly $\Delta_{\rm tr} = 1/3 - P/\varepsilon$ as functions of $n_I$ for the quark-meson model, the pQCD + powers with the perturbative running.
The renormalization scale is fixed to $X=2$ while we examine $\Delta = 0, 200, 300, 400$ MeV.
For large $\Delta$ the trace anomaly can be negative.
}
\label{fig:pQCD_tr_X2}
\end{figure}

\begin{figure}[tbp]
\vspace{-0.3cm}
\centering
\includegraphics[width=1.0\linewidth]{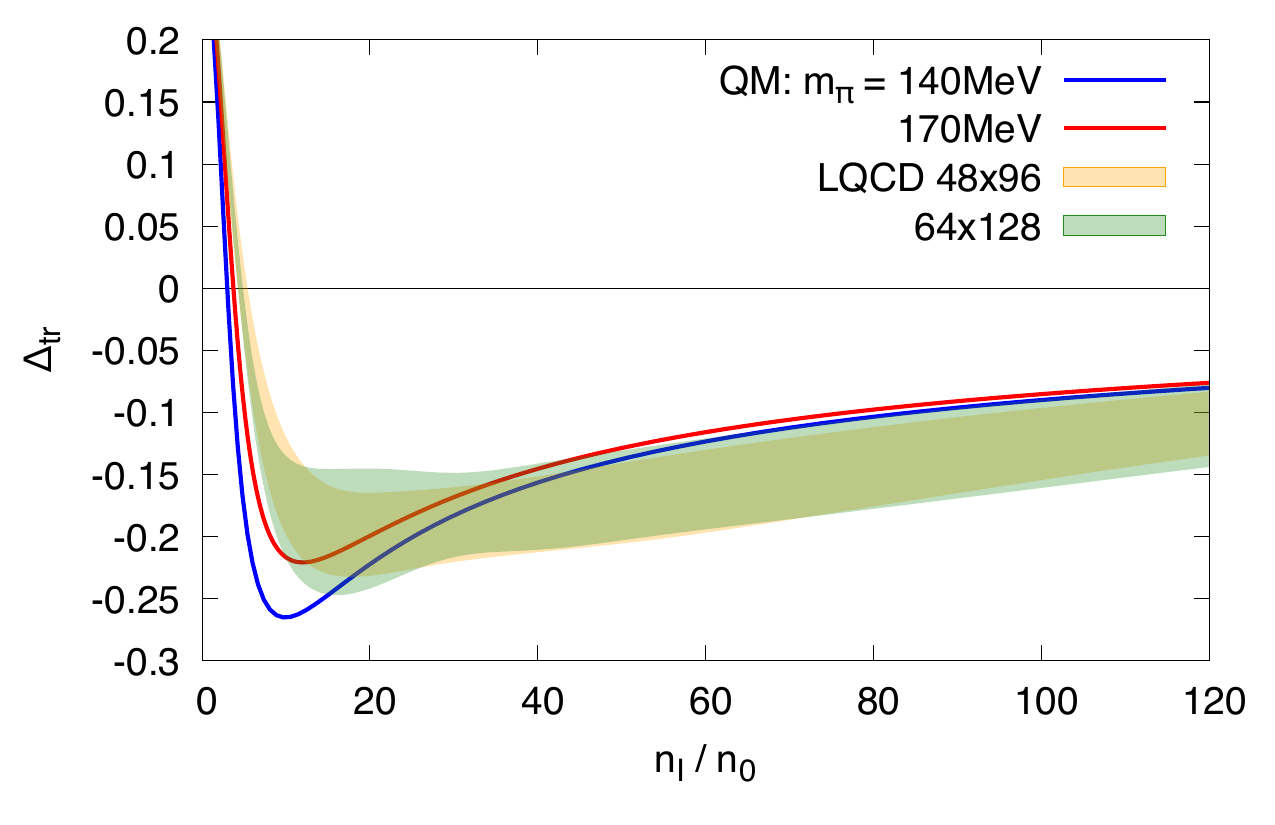}
\caption{ 
Trace anomaly $\Delta_{\rm tr} = 1/3 - P/\varepsilon$ as functions of $\mu_I/m_\pi$ 
for the quark-meson model with $m_\pi =140, 170$ MeV and lattice results of Ref.~\cite{Abbott:2023coj} with $m_\pi \simeq 170$ MeV for different lattice spacing and volume.	
}
\label{fig:trace_anomary_LQCD}
\end{figure}

When we examine the trace anomaly, it is useful to divide it by $3\varepsilon$,
\beq
\Delta_{\rm tr} = \frac{\, 1 \,}{\, 3 \,} - \frac{\, P \,}{\, \varepsilon \,} \,,
\eeq
which should not be confused with the BCS gap $\Delta$.
Shown in Fig.\ref{fig:pQCD_tr_D0} is the $\Delta_{\rm tr}$ as functions of $n_I$
for several calculations, our quark meson model (QM) and pQCD results for the renormalization scales with 
$X=1, 2$, and $4$.
For the pQCD, both the perturbative (dashed) and IR freezing (solid) couplings are examined.
Without power corrections the $\Delta_{\rm tr}$ are all positive.
In Fig.\ref{fig:pQCD_tr_X2}, we fix $X=2$ for these two couplings,
and the dependence of $\Delta_{\rm tr}$ on the power corrections.
The $\Delta = 0, 200, 300$, and $400$ MeV cases are shown.
With power corrections the $\Delta_{\rm tr}$ appears to be negative, as we expected.
Our QM model predicts $\Delta \simeq 300$ MeV and the negative trace anomaly for wide range.
Finally we make a comparison between our QM model results and the lattice results in Ref.~\cite{Abbott:2023coj},
as shown in Fig.~\ref{fig:trace_anomary_LQCD}.
The QM model seems to capture the overall trend of the lattice data.

Since the pQCD corrections and bag constant favor the positive $\Delta_{\rm tr}$,
the negative $\Delta_{\rm tr}$ may be taken as an indicator of the substantial power corrections.

\section{Summary}
\label{sec:summary}

In this work we study the EOS of isospin QCD within a quark-meson model.
The model describes the BEC-BCS crossover of pion condensates.
At tree level pions look elementary, 
but at one-loop they acquire the status of composite particles made of quarks and anti-quarks,
tempering meson fields compared to the tree level amplitudes.
The model is renormalizable and we study its large density behaviors
to study the impacts of non-perturbative physics in the quark matter domain.

Our model exhibits the sound velocity going beyond the conformal limit $c_s^2=1/3$
at $\sim 2n_0$, and making a peak at $\sim 5n_0$, 
at densities substantially smaller than the density for pions to spatially overlap, $\sim 10n_0$.
The quark occupation probability at $p=0$, $f(p=0)$, is $\sim 0.6$ at $2n_0$ and $\sim 0.9$ at $5n_0$. 
The sound velocity peak is located around $5n_0$ where the quark states around $p=0$ are almost fully saturated,
and it makes sense to associate the sound velocity peak with the saturation of quark states.
After the bulk part of the quark Fermi surface is established,
the $c_s^2$ approaches $1/3$ as in the relativistic limit.

Our model shows that $c_s^2$ approaches 1/3 from above,
mainly due to the power corrections, $\sim \mu_I^2 \Delta^2 \sim \mu_I^2 \lqcd^2$.
This sort of terms is not available in pQCD calculations
which predict $c_s^2$ approaching $1/3$ from below.
Which one, perturbative or power corrections, 
dominates in $c_s^2$ around $\sim 40n_0$ is a quantitative question.
The existence of the power corrections
is related to the non-perturbative effects near the quark Fermi surface 
and the structure of the QCD phase diagram.
The question is also related to the sign of the trace anomaly.
The pQCD favors the positive trace anomaly.
If the trace anomaly appears to be negative,
it is strong indication of nontrivial Fermi surface structure.
Lattice results of Ref.~\cite{Abbott:2023coj} seem to support 
the negative trace anomaly in the domain between the BEC and the pQCD domains.
Since the presence of non-perturbative effects in quark matter is a fundamental question,
further clarifications by several lattice calculations with different systematics
are highly desired to establish the findings in Ref.~\cite{Abbott:2023coj}.

The present work left several issues and should be extended to several directions:

i) Our study should be extended to finite temperature (for recent discussions on the quark contributions, see e.g., Refs.~\cite{Blacker:2023onp,Peterson:2023bmr}).
Including thermal effects into quark-meson models is straightforward,
and the results are to be compared with the lattice's.
Whether thermal excitations out of the quark Fermi sea
are confined or deconfined is an important issue
in the context of the quark-hadron-continuity.
As for phenomenological applications to neutron stars and heavy-ion-collisions, 
although several zero temperature EOS have become available since 2012,
finite temperature EOS with the continuity at the level of excitations has not been constructed.
For example, some difficulties have been addressed for the nuclear-2SC continuity in Ref.~\cite{Kojo:2020ztt}.
The magnitude of thermal corrections is much smaller than the cold matter part due to $\sim (T/p_F)^2$ suppression factors,
but it can be important for NSs about to collapse, e.g., those appearing in NS-NS mergers \cite{Huang:2022mqp,Fujimoto:2022xhv,Kedia:2022nns}. 

ii) The estimate of non-perturbative power corrections 
as well as the normalization of EOS (bag constant) at high density should be improved.
Nowadays there has been increasing use of pQCD results 
to constrain the EOS at $\simeq 5$-$40n_0$, with the help of 
general causality and thermodynamic stability conditions (e.g., see Refs.~\cite{Annala:2023cwx,Han:2022rug}).
But as seen in our simple exercise in Sec.\ref{sec:compare_pQCD},
the power corrections of $\sim 10$\% in the overall magnitude
can change the qualitative trend of quantities involving derivatives.
It should be important to see how the power corrections in general 
affect the constraints at $\simeq 5$-$40n_0$.
The present one-loop analyses of quark-meson models should be also improved,
using e.g., the functional renormalization group to include quark and meson fluctuations \cite{Kamikado:2012bt}.

iii) In this work we estimate the density for pion overlap based on
the size of pions in vacuum.
But in medium pions may swell due to the quark exchange among them.
If the effective radii are larger than in the vacuum,
the quark saturation and the overlap of pions can take place
at lower densities than the estimates in this work.
Changes in hadron size may occur already around
nuclear saturation density \cite{Geesaman:1995yd,Saito:2005rv},
as indicated by the comparison of the structure function for
an isolated nucleon and nucleons in nuclei.
It is interesting to test these concepts in isospin QCD
by comparing model predictions with the lattice calculations.

\begin{acknowledgments}
We thank Drs. Brandt and Endrodi for kindly providing us with their lattice data in Ref.~\cite{Brandt:2022hwy},
and Dr. Abbott and his collaborators for their kindness of sending the lattice data in Ref.~\cite{Abbott:2023coj}.
TK thanks Dr. Fujimoto for discussions on $c_s^2$ 
and explanations of his recent works in isospin QCD \cite{Fujimoto:2023unl,Fujimoto:2023mvc}. 
We also thank Dr. Baym for discussions during his visit in GPPU school, 
and Dr. Suenaga for sharing information of his study on hadronic models in isospin QCD \cite{Kawaguchi:2024iaw}.
This work was supported by JSPS KAKENHI Grant No. 23K03377
and by the Graduate Program on Physics for the Universe (GPPU) at Tohoku university.
\end{acknowledgments}

\appendix

\section{Quark propagators}\label{appsec:derivation}

We calculate the mean field quark propagator in the presence of the chiral and pion condensates.
From the propagator one can read off the excitation energy from the pole of the propagator and the occupation probability from the residue of the propagator.

It is convenient to introduce the projection operators for particle and antiparticles,
\beq
	\Lambda_{p,a} = {1\over 2} \pm {\gamma_{j}p^j+M_q \over 2E_D}\gamma_0 \,,
\eeq
which satisfy
\beq
	\Lambda_p + \Lambda_a = 1\,,~ \Lambda_{p,a}\Lambda_{p,a} = \Lambda_{p,a}\,, ~\Lambda_{p,a}\Lambda_{a,p} = 0 \,,
\eeq
as they should.
The propagator of quarks can be written as 
\beq
	{ \rmi \over \Slash{p}+\mu_f\gamma_0-M_q} &=& S_p(p)\gamma_0\Lambda_p+S_a(p)\gamma_0\Lambda_a \,, \notag\\
	S_{p,a}(p) &=& { \rmi \over p_0+\mu_f \mp E_D} \,, 
\eeq
where $S_{p,a}$ is the propagator for a particle and an antiparticle, respectively.

The inverse of the propagator can also be separated by $\Lambda_{p,a}$ as 
\beq
	\Slash{p}+\mu_f\gamma_0-M_q &=& (p_0+\mu_f-E_D)\Lambda_p\gamma_0\notag\\
		&& + (p_0+\mu_f+E_D)\Lambda_a\gamma_0 \,.
\eeq
%
The inverse of the Dirac operator is the quark propagator $S(p)$, and we can write
\beq
	S(p)^{-1} \equiv - \rmi \left(\begin{matrix}\Slash{p} + \mu_u \gamma_{0}-M_q & -i\gamma_5\Delta \\ -i\gamma_5\Delta & \Slash{p} + \mu_d \gamma_{0}-M_q\end{matrix}\right)
\eeq
and consider its inverse.
To simplify the discussion, we introduce the single-particle propagator
\begin{align}
	(G_{u,d}^0)^{-1} = -\rmi \left(\Slash{p} + \mu_{u,d} \gamma_0 -M_q \right) \,,
\end{align}
and write off-diagonal term $\Xi = \gamma_5\Delta$.
Then our propagator must satisfy
\beq
	S(p)^{-1}S(p) = \left(\begin{matrix}(G_u^0)^{-1} & \Xi\\ \Xi & (G_d^0)^{-1}\end{matrix}\right)S(p) = {\bf 1}.
\eeq
Introducing $(G_{u,d})^{-1} \equiv (G_{u,d}^0)^{-1}-\Xi G_{d,u}^0 \Xi$, the propagator $S(p)$ can be written as follows.
\beq
	S(p) = \left(\begin{matrix}G_u & -G_u^0 \Xi G_d\\ -G_d^0 \Xi G_u& G_d\end{matrix}\right).
\eeq
What we are interested in is the diagonal part of $S(p)$ which corresponds to $\langle u\overline{u}\rangle$ and $\langle d\overline{d}\rangle$.
Let us see the detail of $G_{u,d}$.
Its definition is 
\beq
	&&(G_{u,d})^{-1} \equiv (G_{u,d}^0)^{-1}-\Xi G_{d,u}^0 \Xi \notag\\
		&&= - \rmi \left(\Slash{p}+\mu_{u,d}\gamma_0 - M_q \right) -\gamma_5\Delta { \rmi \over\Slash{p} + \mu_{d,u} \gamma_0 - M_q }\gamma_5\Delta \,. \notag\\
\eeq
To calculate the inverse we rewrite this formula using the projection operators.
Performing some calculations we can find
\beq
	\gamma_5\gamma_0 \Lambda_{p,a}\gamma_5 = -\Lambda_{a,p}\gamma_0.
\eeq
From the above, we obtain
\beq
\hspace{-1.0cm}
(G_{u,d})^{-1}
	&=& - \rmi \bigg[
		{\, p_0^2 -(E_D-\mu_{u,d})^2-\Delta^2 \over p_0-\mu_{u,d}+E_D}\Lambda_p\gamma_0 \notag \\
	&& ~+\, {\, p_0^2-(E_D+\mu_{u,d})^2-\Delta^2\over p_0-\mu_{u,d}-E_D}\Lambda_a\gamma_0 
	\bigg].
\eeq
Now we could separate the diagonal elements $(G_{u,d})^{-1}$ by projection operators,
and each part will not be mixed by the inverse operation.

Introducing the excitation energy
\beq
	\xi_{p,a}^f(p) = \sqrt{(E_D\mp\mu_f)^2+\Delta^2},
\eeq
we obtain the propagator
%
\beq
G_f &=& \rmi \bigg[{|u_{p}^f(p)|^2\over \, p_0-\xi_{p}^f(p)}+{|v_{p}^f(p)|^2\over \, p_0+\xi_{p}^f(p)} \bigg]\gamma_0 \Lambda_p  
	\notag \\
       &+& \rmi \bigg[{|u_{a}^f(p)|^2\over \, p_0-\xi_{a}^f(p)}+{|v_{a}^f(p)|^2\over \, p_0+\xi_{a}^f(p)} \bigg]\gamma_0 \Lambda_a \,.
\eeq
Here we have used $\mu_{u,d} = -\mu_{d,u}$.
The residues are
\begin{align}
	|u_{p,a}^f(p)|^2 &= {1\over 2}\left(1+{\pm E_D-\mu_f\over \xi_{p,a}^f}\right)\\
	|v_{p,a}^f(p)|^2 &= {1\over 2}\left(1-{\pm E_D-\mu_f\over \xi_{p,a}^f}\right).
\end{align}
They correspond to the occupation probability and satisfy $|u_p|^2+|v_p|^2=|u_a|^2+|v_a|^2=1$ as expected.
In the main text we use the expressions
\beq
f = f_{u, \bar{d}} &= {1\over 2} \left(1 + { \mu_I - E_D \over E(\mu_I) }\right) = |v_p^u|^2 = |u_a^d|^2 \,, \\
\bar{f} = f_{\bar{u}, d} &= {1\over 2} \left(1 + { \mu_I + E_D \over E(\mu_I) }\right)  = |v_a^u|^2 = |u_p^d|^2\,.
\eeq
%

\section{Full expression of the renormalized one-loop effective potential}\label{appsec:QM_pressure}

In the main text we express the one-loop effective potential using several counter terms.
Rewriting the counter terms using the physical parameters, the final expression turns out to be
%
\begin{widetext}

\begin{align}
V_{\rm 1-loop} 
&= - \frac{1}{\, 4 \,} m_\sigma^2 f_\pi^2 \bigg[
	1 
	+ \frac{4 M_0^2 \Nc}{\, (4\pi)^2 f_\pi^2 \,}  \bigg\{
		- \frac{\, 4 M_0^2 \,}{\, m_\sigma^2 \,} F (m_\sigma^2) 
		+ \frac{\, 4 M_0^2 \,}{\, m_\sigma^2 \,} 
		- ( m_\sigma^2 - 4 M_0^2 ) F'(m_\sigma^2) \bigg\} \bigg] \frac{M_q^2+\Delta^2}{M_0^2} 
\notag\\
&\quad 
+ \frac{3}{\, 4 \,} m_\pi^2 f_\pi^2 \bigg[
	1 
	- \frac{\, 4M_0^2 \Nc \,}{\, (4\pi)^2 f_\pi^2 \,} \bigg\{ - F(m_\pi^2) + F(m_\sigma^2) + (m_\sigma^2 - 4M_0^2 ) F'(m_\sigma^2)\bigg\} \bigg] 
	\frac{\, M_q^2+\Delta^2 \,}{\, M_0^2 \,} 
\notag\\
&\quad - 2 \mu_I^2 f_\pi^2 \bigg[
	1 
	- \frac{\, 4M_0^2 \Nc \,}{\, (4\pi)^2 f_\pi^2 \,} \bigg\{ 
		\ln \frac{\, M_q^2 + \Delta^2 \,}{\, M_0^2 \,} 
		+ F(m_\sigma^2) + ( m_\sigma^2 - 4 M_0^2 ) F' (m_\sigma^2) ) \bigg\} \frac{\Delta^2}{\, M_0^2 \,} 
\notag\\
&\quad 
+ \frac{1}{\, 8 \,} m_\sigma^2 f_\pi^2 \bigg[
	1 
	- \frac{\, 4 M_0^2 \Nc \,}{\, (4\pi)^2 f_\pi^2 \,} \bigg\{ \frac{\, 4M_0^2 \,}{\, m_\sigma^2 \,} \bigg( \ln\frac{\, M_q^2 \,}{\, M_0^2 \,} - \frac{\, 3 \,}{2} \bigg)
	 + \frac{\, 4 M_0^2 \,}{\, m_\sigma^2 \,} F(m_\sigma^2) + (m_\sigma^2 -4 M_0^2 ) F'(m_\sigma^2)) 
 \bigg\} 
 \bigg] \frac{\, (M_q^2+\Delta^2 )^2 \,}{\, M_0^4 \,} 
 \notag\\
&\quad 
- \frac{1}{\, 8 \,} m_\pi^2 f_\pi^2 \bigg[
	1 
	- \frac{\, 4 M_0^2 \Nc \,}{\, (4\pi)^2 f_\pi^2 \,} \bigg\{
	- F(m_\pi^2) + F(m_\sigma^2) + (m_\sigma^2 - 4 M_0^2) F'(m_\sigma^2) \bigg\} \bigg] \frac{\, (M_q^2+\Delta^2)^2 \,}{M_0^4} \notag\\
&\quad 
- m_\pi^2 f_\pi^2 \bigg[
	1 
	- \frac{\, 4 M_0^2 \Nc \,}{\, (4\pi)^2 f_\pi^2 \,} \bigg\{ 
		F(m_\sigma^2) + (m_\sigma^2 - 4 M_0^2 ) F'(m_\sigma^2) \bigg\} \bigg] \frac{\, M_q \,}{\, M_0 \,} 
\notag\\
&\quad
 - 2 \Nc \int_p \bigg[ 
	\sqrt{ \big( E_D + \mu \big)^2+\Delta^2} 
    + \sqrt{ \big( E_D -\mu \big)^2+\Delta^2} 
    - 2 \sqrt{ E_D^2 + \Delta^2 } 
    - \frac{\, \mu_I^2 \Delta^2 \,}{\, (E_D^2 + \Delta^2)^{3/2} \,}
    \bigg]
 \,.
 \label{eq:eff_pot_hadronic}
\end{align}
\end{widetext}
The function $F(p^2)$ is given by
\beq
	F(p^2) = 2-2r\arctan \left({1\over r}\right)\\
	p^2F'(p^2) = {r^2+1 \over r}\arctan\left({1\over r}\right)-1
\eeq
where $r=\sqrt{4{M_0}^2/p^2-1}$.
This parametrization suggests that the parameters $m_\pi$ and $m_\sigma$ are restricted to $m_\pi,m_\sigma < 2{M_0}$ to make $r$ real.

\bibliography{ref}

\end{document}